\documentclass[letterpaper, reprint, showkeys, nofootinbib, prl, aps, superscriptaddress, longbibliography, nobalancelastpage]{revtex4-1}

\usepackage{color}

\usepackage[pdftex, breaklinks=true]{hyperref}
\hypersetup{
    pdffitwindow=true,     
    pdfstartview={FitH},    
    pdfnewwindow=true,      
    colorlinks=true,       
    linkcolor=red,      
    citecolor=blue,        
    urlcolor=blue           
}
\usepackage[utf8]{inputenc}
\DeclareSymbolFont{slenderlargesymbols}{OMX}{ccex}{m}{n}
\DeclareMathSymbol{\prod}{\mathop}{slenderlargesymbols}{"51}
\usepackage{amsfonts, amssymb, amsmath}
\usepackage{amsthm}
\usepackage{mathtools}
\usepackage{braket} 
\usepackage{multirow}
\usepackage[T1]{fontenc}
\usepackage{bbm}
\usepackage{float}
\usepackage{caption}
\usepackage{subcaption}

\newcommand{\ketbra}[2]{|#1\rangle \langle#2|}

\newtheoremstyle{cited}%
  {3pt}
  {3pt}
  {\itshape}
  {}
  {\bfseries}
  {.}
  {.5em}
  {\thmname{#1} \thmnumber{#2} \thmnote{\normalfont#3}}
\theoremstyle{cited}

\makeatletter

\newcommand\k@t[1]{{|{#1}\rangle}}
\makeatother
\DeclarePairedDelimiter{\abs}{\lvert}{\rvert}

\DeclareMathOperator{\Tr}{\text{Tr}}

\usepackage{xcolor}

\usepackage[normalem]{ulem}

\usepackage{tikz}
\usetikzlibrary{decorations.pathreplacing}
\makeatletter
\def\underbrace#1{%
   \@ifnextchar_{\tikz@@underbrace{#1}}{\tikz@@underbrace{#1}_{}}}
\def\tikz@@underbrace#1_#2{%
   \tikz[baseline=(a.base)] {\node[inner sep=2] (a) {\(#1\)};
   \draw[line cap=round,decorate,decoration={brace,amplitude=4pt}]
     (a.south east) -- node[pos=0.5,below,inner sep=7pt] {\(\scriptstyle #2\)} (a.south west);}}
 
\def\overbrace#1{%
   \@ifnextchar^{\tikz@@overbrace{#1}}{\tikz@@overbrace{#1}^{}}}
\def\tikz@@overbrace#1^#2{%
   \tikz[baseline=(a.base)] {\node[inner sep=2] (a) {\(#1\)};
   \draw[line cap=round,decorate,decoration={brace,amplitude=4pt}]
     (a.north west) -- node[pos=0.5,above,inner sep=7pt] {\(\scriptstyle #2\)} (a.north east);}}
\makeatother

\usepackage{siunitx}

\makeatletter 
    
\renewcommand\onecolumngrid{
\do@columngrid{one}{\@ne}%
\def\set@footnotewidth{\onecolumngrid}
\def\footnoterule{\kern-6pt\hrule width 1.5in\kern6pt}%
}

\renewcommand\twocolumngrid{
        \def\footnoterule{
        \dimen@\skip\footins\divide\dimen@\thr@@
        \kern-\dimen@\hrule width.5in\kern\dimen@}
        \do@columngrid{mlt}{\tw@}
}%

\makeatother    

\begin{document}

\begin{abstract}
  Tensor network quantum machine learning (QML) models are promising applications on near-term quantum hardware. While decoherence of qubits is expected to decrease the performance of QML models, it is unclear to what extent the diminished performance can be compensated for by adding ancillas to the models and accordingly increasing the virtual bond dimension of the models. We investigate here the competition between decoherence and adding ancillas on the classification performance of two models, with an analysis of the decoherence effect from the perspective of regression. We present numerical evidence that the fully-decohered unitary tree tensor network (TTN) with two ancillas performs at least as well as the non-decohered unitary TTN, suggesting that it is beneficial to add at least two ancillas to the unitary TTN regardless of the amount of decoherence may be consequently introduced.
\end{abstract}

\title{Decohering Tensor Network Quantum Machine Learning Models}

\author{Haoran Liao}
\email[E-mail: ]{haoran.liao@berkeley.edu}
\affiliation{Department of Physics, University of California, Berkeley, CA 94720, USA}
\affiliation{Berkeley Quantum Information and Computation Center, University of California, Berkeley, CA 94720, USA}
\author{Ian Convy}
\affiliation{Department of Chemistry, University of California, Berkeley, CA 94720, USA}
\affiliation{Berkeley Quantum Information and Computation Center, University of California, Berkeley, CA 94720, USA}
\author{Zhibo Yang}
\affiliation{Department of Chemistry, University of California, Berkeley, CA 94720, USA}
\affiliation{Berkeley Quantum Information and Computation Center, University of California, Berkeley, CA 94720, USA}
\author{K. Birgitta Whaley}
\affiliation{Department of Chemistry, University of California, Berkeley, CA 94720, USA}
\affiliation{Berkeley Quantum Information and Computation Center, University of California, Berkeley, CA 94720, USA}

\date{\today}
\maketitle

\section{Introduction}
 Tensor networks (TNs) are compact data structures engineered to efficiently approximate certain classes of quantum states used in the study of quantum many-body systems. Many tensor network topologies are designed to represent the low-energy states of physically realistic systems by capturing certain entanglement entropy and correlation scalings of the state generated by the network \cite{Evenbly_2011, Eisert_2013, Convy_2022, Cirac_2021}. Some tensor networks allow for interpretations of coarse-grained states at increasing levels of the network as a renormalization group or scale transformation that retains information necessary to understand the physics on longer length scales \cite{Evenbly_Vidal_2009, Hand-waving}. This motivates the usage of such networks to perform discriminative tasks, in a manner similar to classical machine learning (ML) using neural networks with layers like convolution and pooling that perform sequential feature abstraction to reduce the dimension and to obtain a hierarchical representation of the data \cite{levine_2018, Cohen_2016}. 
In addition to applying TNs such as the tree tensor network (TTN) \cite{ttn_origin} and the multiscale entanglement renormalization ansatz (MERA) \cite{Vidal_2007} for quantum-inspired tensor network ML algorithms \cite{Stoudenmire_2018, Reyes_Stoudenmire_2021, Wall_2021}, there have been efforts to variationally train the generic unitary nodes in TNs to perform quantum machine learning (QML) on data-encoded qubits. The unitary TTN \cite{Grant_2018, Huggins_2019} and MERA \cite{Grant_2018, Cong2019} have been explored for this purpose mindful of feasible implementations, such as normalized input states, on a quantum computer. 

Tensor network QML models are linear classifiers on a feature space whose dimension grows exponentially in the number of data qubits and where the feature map is non-linear. Such models employ fully-parametrized unitary tensor nodes that form a rich subset of larger unitaries with respect to all input and output qubits upon tensor contractions. They provide circuit variational ansatze more general than those with common parametrized gate sets \cite{Mitarai_2018, Benedetti_2019, Havlicek_2019}, although their compilations into hardware-dependent native gates are more costly because of the need to compile generic unitaries.

In this work, we focus on discriminative QML. We investigate and numerically quantify the competing effect between decoherence and increasing bond dimension of two common tensor network QML models, namely the unitary TTN and the MERA. By removing the off-diagonal elements, i.e., the coherence, from the density matrix of a quantum state, we reduce its representation down to a classical probability distribution over a given basis. The evolution through the unitary matrices at every layer of the model, together with the full dephasing of the density matrix at input and output, then becomes successive Bayesian updates of classical probability distributions, thus removing the quantumness of the model. This process can occur between any two layers of the unitary TTN or the MERA, and should in principle reduce the amount of information or representative flexibility available to the classification algorithm. However, as we add and increase the number of ancillas and accordingly increase the virtual bond dimension of the tensor networks, this diminished expressiveness may be compensated by the increased dimension of the classical probability distributions and their conditionals, manifested in the increasing number of diagonals intermediate within the network, as well as by the increased sized of the stochastic matrices encapsulated by the corresponding Bayesian networks in the fully-dephased limit. The possibility that an increased bond dimension fully compensates for the decoherence of the network would indicate that the role of coherence in QML is not essential and it offers no unique advantage, whereas a partial compensation provides insights into the trade-off between adding ancillas and increasing the level of decoherence in affecting the network performance, and therefore offers guidance in determining the number of noisy ancillas to be included in NISQ-era~\cite{preskill_quantum_2018} implementations.

The remainder of the paper is structured as follows. Sec.~\ref{sec:prelim} explains two tensor network QML models, the unitary TTN and the MERA. Sec.~\ref{sec:dephasing} reviews the dephasing effect on quantum states and shows its effect on the models from the perspective of regression. In Sec.~\ref{sec:ancillas} we explain the scheme in which ancillas are added to the networks and the growth of the virtual bond dimensions of the networks. Sec.~\ref{sec:related_work} summarizes related work to unify fully-dephased tensor networks into probabilistic graphical models. In Sec.~\ref{sec:numerical_experiments} we numerically experiment on natural images to show the competing effect between decoherence and adding ancillas while accordingly increasing the virtual bond dimension of the network. Sec.~\ref{sec:discussion} summarizes and discusses the conclusions. In App.~\ref{sec:fully-dephase_tn}, a formal mathematical treatment to connect the fully-dephased tensor networks to classical Bayesian networks is presented.

\section{Preliminaries}\label{sec:prelim}
\subsection{Tensor Network QML Models}
\subsubsection{Unitary TTN}
\begin{figure*}
 \centering
\includegraphics[scale=0.3]{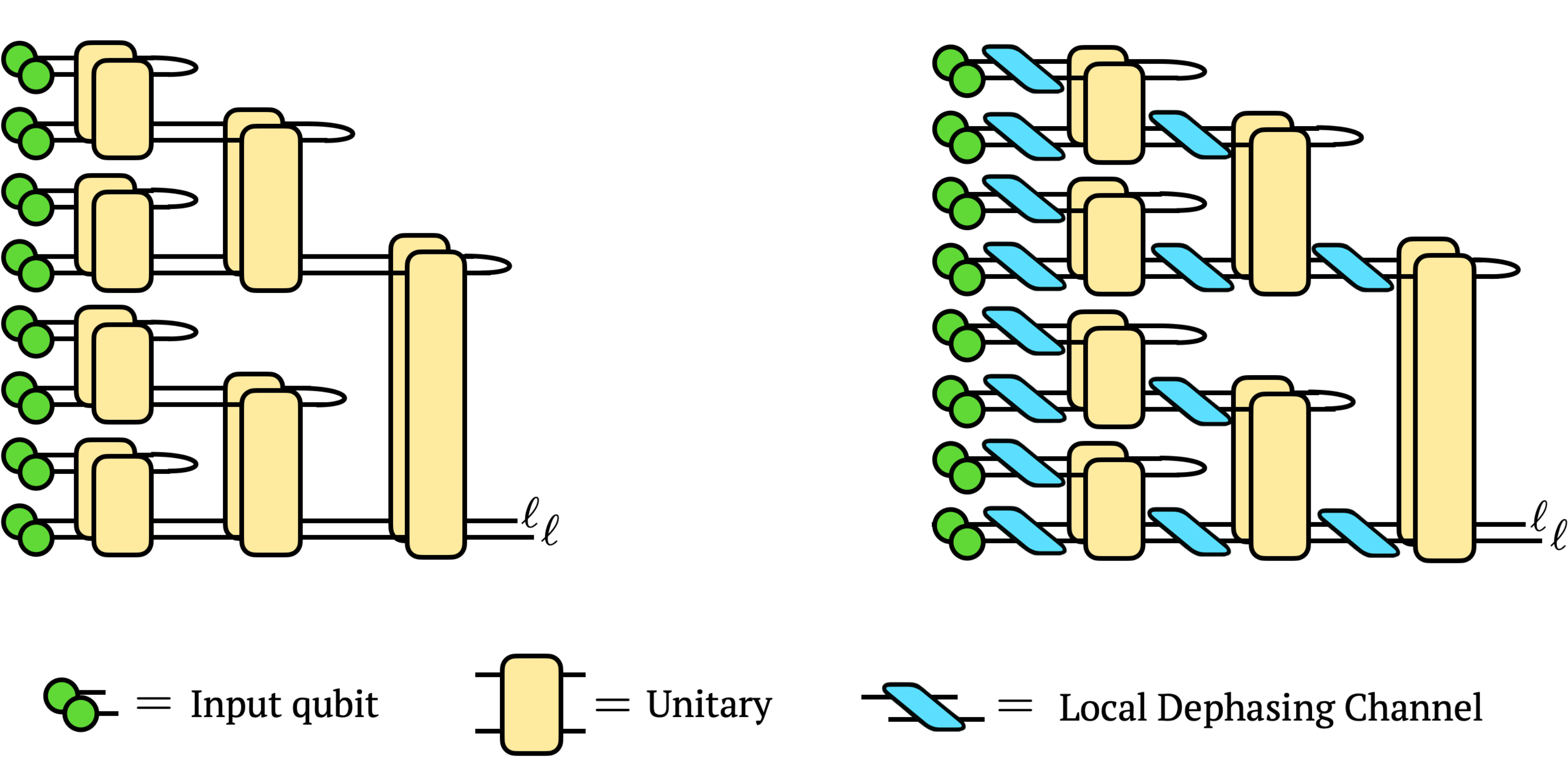}
\captionsetup{justification=raggedright, singlelinecheck=false}
 \caption{Left: A unitary TTN on eight input features encoded in the density matrices $\rho_{\text{in}}$'s forming the data layer, where the basis state $\ell$ is measured at the output of the root node. Right: Dephasing the unitary TTN is to insert dephasing channels with a dephasing rate $p$, assumed to be uniform across all, into the network between every layer.}
 \label{fig:ttn}
\end{figure*}

Unitary TTN is a classically tractable realization of tensor network QML models, with a topology that can be interpreted as a local coarse-graining transformation that keeps the most relevant degrees of freedom, in a sense that the information contained within each subtree is separated from those contained outside of the subtree. We focus on 1D binary trees. A generic binary TTN consists of $\log(m)$ layers of nodes where $m$ is the number of input features, plus a layer of data qubits appended to the leaf level of the tree. A diagram of the unitary TTN is shown in Fig.~\ref{fig:ttn} (left). Every node in a unitary TTN is forced to be a unitary matrix with respect to its input and output Hilbert spaces. Each unitary tensor entangles a pair of inputs from the previous layer. At each layer, one of the two output qubits is unobserved and also not further operated on, while the other output qubit is evolved by a node at the next layer. If the classification is binary, at the output of the last layer, namely the root node, only one qubit is measured. Accumulation of measurement statistics then reveals the confidence in predicting the binary labels associated with the measurement basis. After variationally learning the weights in the unitary nodes, we recover a quantum channel such that the information contained in the output qubits of each layer can be viewed as a coarse-grained representation of that in the input qubits, which sequentially extracts useful features of the data encoded in the data qubits. A dephased unitary TTN has local \textit{dephasing channels} inserted between any two layers of the network, as depicted in Fig.~\ref{fig:ttn}~(right).

\subsubsection{MERA}
In tensor network QML, the MERA topology overcomes the drawback of local coarse-graining in unitary TTN by adding disentanglers $U$, which are unitaries, to connect neighboring subtrees. Its subsequent decimation of the Hilbert space by a MERA is achieved by isometries $V$ that obey the isometric condition only in the reverse coarse-graining direction, i.e., $V^\dagger V=I'$ but $VV^\dagger\neq I$. From the perspective of discriminative QML, these unitaries correlate information from states in neighboring subtrees. We thus refer to these unitaries as entanglers.

\begin{figure*}
 \centering
\includegraphics[scale=0.35]{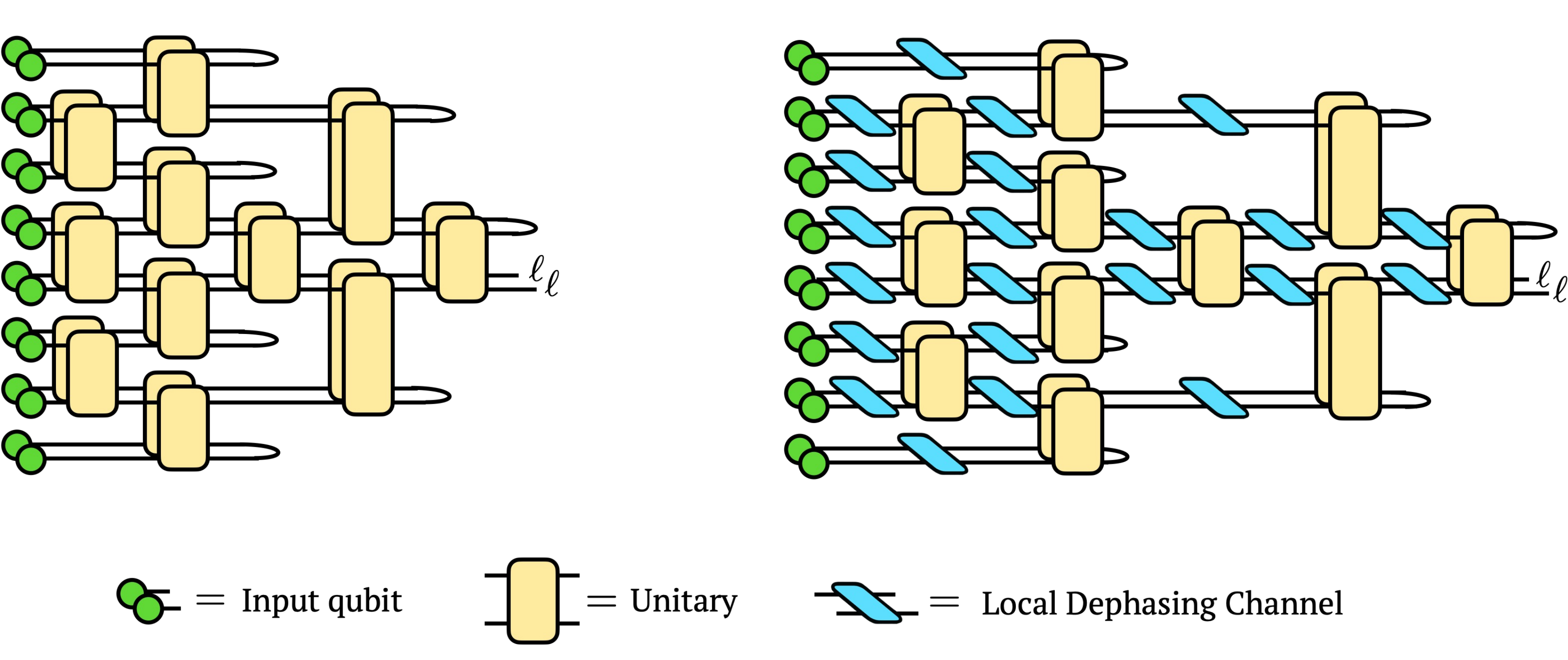}
\captionsetup{justification=raggedright, singlelinecheck=false}
  \caption{Left: A MERA on eight input features encoded in the $\rho_{\text{in}}$'s forming the data layer, where the basis state $\ell$ is measured at the output of the root node. Right: Dephasing the MERA is to insert dephasing channels with a dephasing rate $p$, assumed to be uniform across all, into the network between every layer.}
 \label{fig:mera}
\end{figure*}

By the design of MERA~\cite{Vidal_2007}, the adjoint of an isometry, namely an isometry viewed in the coarse-graining direction in QML, can be naively achieved by measuring one of the two output qubits in the computational basis and post-selecting runs with measurements yielding $\ket{0}$. However, this way of decimating the Hilbert space is generally prohibitive, given the vanishing probability of sampling a bit string of all output qubits with most of them in $\ket{0}$. Hence, operationally an isometry is replaced by a unitary node, half of whose output qubits are partially traced over, which is the same as a unitary node in the TTN. The MERA can now be understood as a unitary TTN with extra entanglers inserted before every tree layer except the root layer, such that they entangle states in neighboring subtrees, as shown in Fig.~\ref{fig:mera} (left). Its dephased version is similar to the dephased unitary TTN, as depicted in Fig.~\ref{fig:mera} (right).

\section{Dephasing}\label{sec:dephasing}
\subsection{Dephasing Qubits after Unitary Evolution}\label{sec:dephase_basics}
A dephasing channel with a rate $p\in(0, 1]$ on a qubit is obtained by tracing out the environment after the environment scatters off of the qubit with some probability $p$. We denote the dephasing channel on a qubit with a dephasing rate $p$ as $\mathcal{E}$, such that 
\begin{equation}\label{eq:dephase}
\begin{split}
    &\mathcal{E}[\rho]=(1-\frac{1}{2}p)\rho+\frac{1}{2}p\sigma_3\rho\sigma_3\\
    &=\sum_{ij}(1-p)^{1-\delta_{ij}}\braket{i|\rho|j}\ketbra{i}{j}=\sum_{ij}(1-p)^{1-\delta_{ij}}\rho_{ij}\ketbra{i}{j},
\end{split}
\end{equation}
where the summation goes from $0$ to $1$ for every index hereafter unless specified otherwise, whose effect is to damp the off-diagonal entries of the density matrix by $(1-p)$. The operator-sum representation of $\mathcal{E}[\rho]$ can be written as with the two Kraus operators\footnote{A more commonly-used, but less computationally efficient in terms of Eq.~\eqref{eq:multi_qubits_dephase}, representation uses three Kraus operators: $K_0=\sqrt{1-p}I$ and $K_{1/2}=\frac{\sqrt{p}}{2}(I\pm\sigma_3)$ such that $\mathcal{E}[\rho]=\sum_{i=0}^2 K_i\rho K_i^\dagger$ and $\sum_{i=0}^2 K_i^\dagger K_i=I$.},
\begin{equation}
    K_0=\sqrt{1-\frac{p}{2}}I,\quad K_1=\sqrt{\frac{p}{2}}\sigma_3,
\end{equation}
defined such that $\mathcal{E}[\rho]=\sum_{i}K_i\rho K_i^\dagger$ and $\sum_{i} K_i^\dagger K_i=I$. Assuming local dephasing on each qubit, the dephasing channel on the density matrix $\rho$ of $m$ qubits, entangled or not, is given by
\begin{equation}\label{eq:multi_qubits_dephase}
    \mathcal{E}[\rho]=\sum_{i_1,\dots, i_m}\left(\bigotimes_{n=1}^m K_{i_n}\right)\rho \left(\bigotimes_{n=1}^m K_{i_n}^\dagger\right).
\end{equation}

If we allow a generic unitary $U$ to act on $\mathcal{E}[\rho]$ for a single qubit,
we have the purity of the resultant state given by
\begin{equation}
\begin{split}
        \Tr&\left[\left(U\mathcal{E}[\rho]U^\dagger\right)^2\right]=\Tr\left[ \left( \left(1-\frac{p}{2}\right)\rho+\frac{p}{2}\sigma_3\rho\sigma_3 \right)^2 \right]\\
        &=\Tr\left(\rho^2\right)-4p\rho_{01}^2\left(1-\frac{p}{2}\right)\leq\Tr\left(\rho^2\right),
\end{split}
\end{equation}
where we used Eq.~\eqref{eq:dephase} in the first line. Therefore, in a given basis, successive applications of a dephasing channel and generic unitary evolution decrease the purity of any input quantum state, until the state becomes maximally mixed\footnote{Unitary evolution on the $d$-dimensional maximally mixed states, which are the only rotationally invariant states, does not produce coherence.}. Successively applying the dephasing channel alone decreases the purity of the state until it becomes fully decohered, namely diagonal in its density operator in a given basis. It is thus a process in which quantum information of the input is irreversibly and gradually (for $p<1$) lost to the environment until the state becomes completely describable by a discrete classical probability distribution.

\subsection{Dephasing Product-state Encoded Input Qubits}\label{sec:dephase_input}

When inputting data into a tensor network, it is common to featurize each sample into a product state, or a rank-one tensor. The density matrix of such a state with $m$ features is given by $\rho = \bigotimes^m_{n=1} \ketbra{f^{(n)}}{f^{(n)}}=\bigotimes^m_{n=1}\rho^{(n)}$,
where $\ket{f^{(n)}}$ is a state of dimension $d$ that encodes the $n$th feature. Assuming local dephasing on each data qubit, it is expected that the product state density matrix after dephasing is the product state of the dephased component density matrix, i.e., $\mathcal{E}[\rho]=(\bigotimes_{n=1}^m\mathcal{E}^{(n)})[\bigotimes^m_{n=1}\rho^{(n)}]=\bigotimes^N_{n=1}\mathcal{E}^{(n)}[\rho^{(n)}]$.

In the context of our tensor network classifier, the effect of dephasing can be seen by considering just a single feature. If we normalize this feature such that its value is $x^{(n)} \in [0, 1]$, then we can utilize the commonly-used qubit encoding \cite{stoudenmire_qubit_encoding, Larose_2020, Liao_2021} to encode this classical feature into a qubit as
\begin{equation}\label{eq:qubit_encoding}
    \ket{f^{(n)}} = 
    \begin{bmatrix}
        \sin\left(\frac{\pi}{2}x^{(n)}\right) \\
        \cos\left(\frac{\pi}{2}x^{(n)}\right)
    \end{bmatrix},
\end{equation}
respectively. A notable property of these encodings is that the elements of $\ket{f^{(n)}}$ are always positive, so there is a one-to-one mapping between $\abs{\braket{i^{(n)}|f^{(n)}}}^2$ and $\braket{i^{(n)}|f^{(n)}}$ for all $i^{(n)}$. This means that every element of $\rho^{(n)} = \ketbra{f^{(n)}}{f^{(n)}} \equiv \rho$ can be written as a function of probabilities $\lambda_0^{(n)}\equiv\lambda_0$ and $\lambda_1^{(n)}\equiv\lambda_1$, where
\begin{equation} \label{eq:lambda}
    \rho_{00}=\lambda_0, \quad \rho_{01} = \rho_{10} = \sqrt{\lambda_0\lambda_1}, \quad \rho_{11} = \lambda_1.
\end{equation}
Using Eq.~\eqref{eq:fully_dep_after_U}, we get
\begin{align}
    \lambda'_0 &= |U_{00}|^2\lambda_0 + |U_{01}|^2\lambda_1 + 2\sqrt{\lambda_0\lambda_1}\Re(U_{00}U_{01})
    \\
    \lambda'_1 &= |U_{11}|^2\lambda_1 + |U_{10}|^2\lambda_0 + 2\sqrt{\lambda_0\lambda_1}\Re(U_{10}U_{11}),
\end{align}
where it is clear that the new probabilities $\lambda'_i$ are non-linear functions of the old probabilities $\lambda_j$. Specifically, there is a dependence on $\sqrt{\lambda_0\lambda_1}$. Such non-linear functions cannot be generated by a stochastic matrix acting on $\text{diag}(\rho^{(n)})$, since the off-diagonal $\sqrt{\lambda_0\lambda_1}$ terms will be set to zero. By fully dephasing the input state before acting the unitary, the fully-dephased output is less expressive in the sense that we lose the regressor $\sqrt{\lambda_0\lambda_1}$. But knowing the relative phase of the encoding, this lost regressor does not contain any extra information than the regressors $\lambda_0$ and $\lambda_1$, so in that sense the information content of the encoding is unaffected by the dephasing.

\subsection{Impact on Regressors by Dephasing}
To understand the dephasing effect on the linear regression induced by the unitary TTN network topology, it is illuminating to study the evolution of $\Tr_A(U\mathcal{E}[\rho]U^\dagger)$ which is undertaken by a unitary node acting on a pair of dephased input qubits followed by a partial tracing over one of the output qubits. 
The diagonals of the output density matrix before partial tracing, i.e., the diagonals of $U\mathcal{E}[\rho]U^\dagger$, are
\begin{equation}\label{eq:dephased_reg_bf_tracing}
\begin{split}
    \rho'_{ii}&=
    \abs{U_{i0}}^2\rho_{00}+\abs{U_{i1}}^2\rho_{11}+\abs{U_{i2}}^2\rho_{22}+\abs{U_{i3}}^2\rho_{33}+\\
    &\quad 2(1-p)\left[\Re(U_{i1}U_{i0}^*\rho_{10})+\Re(U_{i2}U_{i0}^*\rho_{20})+\right.\\
    &\left. \quad\quad\quad\quad\quad \Re(U_{i3}U_{i1}^*\rho_{31})+\Re(U_{i3}U_{i2}^*\rho_{32})
    \right]+\\
    &\quad 2(1-p)^2\left[\Re(U_{i3}U_{i0}^*\rho_{30})+\Re(U_{i2}U_{i1}^*\rho_{21})\right],
\end{split}
\end{equation}
for $i\in\{0,1,2,3\}$, where every diagonal term is a linear regression on all elements of input $\rho$ with regression coefficients set by the unitary matrix elements $U_{ik}, k\in\{0,1,2,3\}$. We note that terms such as the $\Re(U_{i1}U_{i0}^*\rho_{10}) = U_{i0}U_{i1}^*\rho_{01}+U_{i1}U_{i0}^*\rho_{10}$ are each composed of two regressors. In particular, the dephasing suppresses some of the regressors by a factor of $(1-p)$ or $(1-p)^2$. Since the norm of each element in $U$ and $U^\dagger$ is upper bounded by one, the norm of the regression coefficients is suppressed by these factors induced by dephasing. The suppression is stronger by a factor of $(1-p)^2$ for regressors that are anti-diagonals of the input density matrix, i.e., $\rho_{30}$ and $\rho_{21}$. While the regression described above is to obtain the diagonals of the output density matrix, the regression to obtain off-diagonals of the output density matrix has a similar pattern of suppression of certain regressors.

This suppression of regression coefficients is carried over to the reduced density matrix, which can be written as
\begin{equation}\label{eq:dephased_regression}
    \Tr_2(\rho')=
  \begin{bmatrix}
    \rho'_{00}+\rho'_{11} & \rho'_{02}+\rho'_{13}\\
    \rho'_{20}+\rho'_{31} & \rho'_{22}+\rho'_{33}\\
  \end{bmatrix}.
\end{equation}

When the input pair of qubits $\rho$ is a product state of two data qubits, we have
\begin{equation}\label{eq:two_qubit}
\begin{split}
  \rho = \rho^{(1)}\otimes\rho^{(2)} 
  &\equiv \begin{bmatrix}
    \lambda_0 & \sqrt{\lambda_0\lambda_1}\\
    \sqrt{\lambda_0\lambda_1} & \lambda_1\\
  \end{bmatrix}
  \otimes
  \begin{bmatrix}
  \mu_0 & \sqrt{\mu_0\mu_1}\\
  \sqrt{\mu_0\mu_1} & \mu_1\\
  \end{bmatrix},\\
 \end{split}
\end{equation}
where the $\lambda$'s and $\mu$'s are defined like Eq.~\eqref{eq:lambda} for the two data qubits $\rho^{(1)}$ and $\rho^{(2)}$. Substituting Eq.~\eqref{eq:two_qubit} into Eq.~\eqref{eq:dephased_reg_bf_tracing}~and~\eqref{eq:dephased_regression}, we see that all regressors containing $\sqrt{\mu_0\mu_1}$ or $\sqrt{\lambda_0\lambda_1}$ are suppressed by a factor of $(1-p)$ after the first-layer unitary, while the regressor $\sqrt{\lambda_0\lambda_1\mu_0\mu_1}$ is suppressed by a factor of $(1-p)^2$. The output density matrix elements then become the regressors for regressions performed by subsequent upper layers, as follows.

For unitary TTN without \textit{ancillas}, Eq.~\eqref{eq:dephased_reg_bf_tracing}~and~ \eqref{eq:dephased_regression} are carried over to the output of every layer of the network, since there is no entanglement in the input pair of qubits. However, at the upper layers, the regression onto the output density matrix element has regressors already composed of terms that were suppressed in previous layers, as described above for $\rho\rightarrow\rho'$. Viewing the regressors at the input of the last layer, the suppression on most of them by some power of $(1-p)$ resembles the concept of regularization in regressions but does not involve a penalty term on the coefficient norm in the loss function.

In cases where there can be entanglement in each of the input qubits, such as the intermediate layers in a MERA or in a unitary TTN with ancillas, the pattern of suppressing certain regressors is similar, where the coherence of the input is suppressed by some power of $(1-p)$. In particular, the regressors on the anti-diagonals are most strongly suppressed by a factor of $(1-p)^{m}$ where $m$ is the number of input qubits.

\subsection{Fully-dephased Unitary Tensor Networks}
When the network is fully-dephased at every layer, all of the off-diagonal regressors are removed. Each diagonal term of the output density matrix then becomes a regression on only the diagonals of the input density matrix. In App.~\ref{sec:dephase_reduced_rho}, we show that in this situation each node of the unitary tensor network $U_{ij}$ reduces to a unitary-stochastic matrix $M_{ij} \equiv |U_{ij}|^2$. When the output of the unitary node is partially traced over, the overall operation is equivalent to a singly stochastic matrix $S_{i_B j}\equiv \sum_{i_A}\abs{U_{i_Ai_Bj}}^2$, where $i_A$ enumerates the traced-over part of the system. The tensor network QML model then reduces to a classical Bayesian network (see App.~\ref{sec:bayes_net}) with the joint probability factorization Eq.~\eqref{eq:bayesian_network_factorization} presented in App.~\ref{sec:dephasing_uni_ttn}~and~\ref{sec:dephasing_mera}.

\section{Adding Ancillas and Increasing the Virtual Bond Dimension}\label{sec:ancillas}

The Stinespring's dilation theorem \cite{Kretschmann_2008, watrous_2018} states that any quantum channel or completely positive and trace-preserving (CPTP) map $\Lambda: \mathcal{B}(\mathcal{H}_A)\rightarrow\mathcal{B}(\mathcal{H}_B)$\footnote{We denote the convex set of positive-semidefinite linear operators with unit trace, namely the set of density operators, on a complex Hilbert space $\mathcal{H}$ (thus Hermitian and bounded) as $\mathcal{B}(\mathcal{H})$.} over finite-dimensional Hilbert spaces $\mathcal{H}_A$ and $\mathcal{H}_B$ is equivalent to a unitary operation on a higher dimensional Hilbert space $\mathcal{H}_B\otimes \mathcal{H}_E$, where $\mathcal{H}_E$ is also finite-dimensional, followed by a partial tracing over $\mathcal{H}_E$. A motivating example demonstrating directly that ancillas are necessary to allow the evolution of fully-dephased input induced by a generic unitary to be as expressive as that induced by a singly stochastic matrix is presented in App.~\ref{sec:stinespring}. In particular, the dimension of the ancillary system $\mathcal{H}_E$ can be chosen such that $\dim(\mathcal{H}_E)\leq\dim(\mathcal{H}_A)\dim(\mathcal{H}_B)$ for any $\Lambda$\footnote{In the Stinespring's representation of such a CPTP map $\Lambda$, there exists an isometry $V: \mathcal{B}(\mathcal{H}_A)\rightarrow\mathcal{B}(\mathcal{H}_B\otimes\mathcal{H}_E)$ such that $\Lambda(\rho)=\Tr_E(V\rho V^\dagger), \forall \rho\in\mathcal{B}(\mathcal{H}_A)$.} \cite{Kretschmann_2008}. In terms of qubits, the theorem implies that there need to be at least $2n_o$ ancilla qubits to achieve an arbitrary quantum channel between $n_i$ input qubits and $n_o$ output qubits. This is because the total combined number of $n_i$ input qubits and $n_a$ ancilla qubits should equal the total combined number of $n_o$ output qubits and the qubits that are traced out as environment degrees of freedom. Using Stinespring's dilation theorem, we can show $2^{n_i+n_a-n_o}\leq2^{n_i}2^{n_o}$ which leads to $n_a\leq 2n_o$.

In the scheme of adding ancillas per node in a unitary TTN, every node requires then in principle at least two ancilla qubits to achieve an arbitrary quantum channel, because there are two input qubits coming from the previous layer and one output qubit passing to the next layer.
\begin{figure}[H]
 \centering
\includegraphics[scale=0.25]{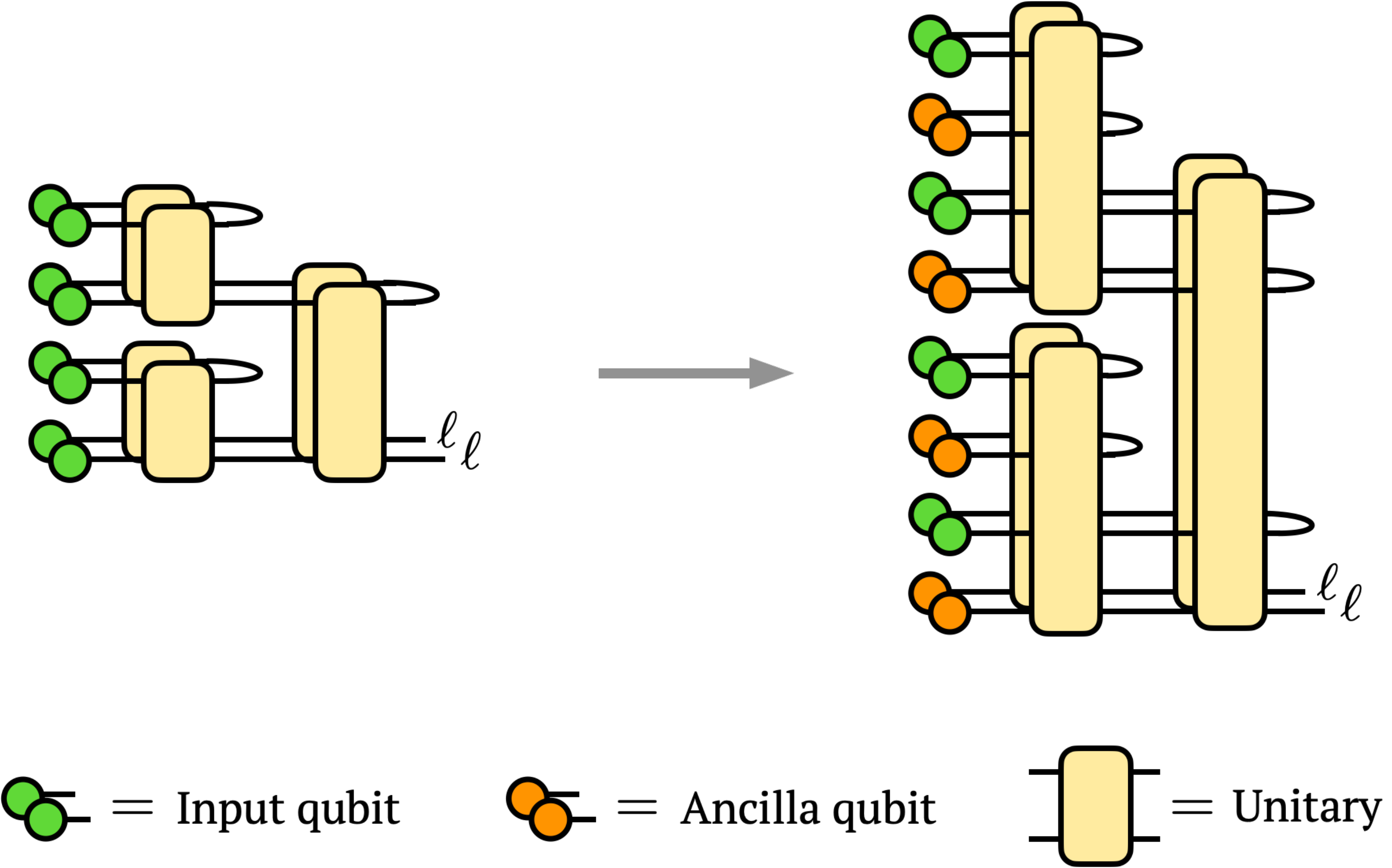}
 \caption{Adding one ancilla qubit, initialized to a fixed basis state, per data qubit to a unitary TTN classifying four features, with a corresponding virtual bond dimension increased to four. Only one output qubit is measured in the basis state $\ell$ regardless of the number of ancillas added per data qubit. We always decimate the Hilbert space by half between consecutive layers of unitary nodes.}
 \label{fig:ancilla_ttn}
\end{figure}

However, in practice, we have found it more expressive to instead add ancillas to the data qubits and to trace out half of all output qubits per node before contracting with the node at the next layer. We call this the ancilla-per-data-qubit scheme. This scheme is able to achieve superior classification performance in the numerical experiment tasks that we conducted compared to the ancilla-per-unitary-node scheme described above (see details in App.~\ref{app:adding_ancillas}), despite the fact that the two schemes share the same number of trainable parameters when adding the same number of ancillas. A diagram of this ancilla scheme is shown in Fig.~\ref{fig:ancilla_ttn}. This scheme effectively increases the virtual bond dimension of the network, which means that the network can represent a larger subset of unitaries on all input qubits.

Although the ancilla-per-data-qubit scheme achieves superior classification performance, it never produces arbitrary quantum channels at each node. To see this, for any unitary node in the first layer, the number of input qubits is $n_i=2$, that of ancillas is $n_a=n_ik=2k$ where $k\in\mathbbm{Z}$ is the number of ancillas per data qubit, and that of output qubits passing to the next layer is $n_o=1+k$ such that $n_a<2n_o, \forall a\in\mathbbm{Z}$. As a result, the channels achievable via the first layer of unitaries constitute only a subset of all possible channels between its input and output density matrices. For any unitary node in subsequent layers, there are no longer any ancillas, whereas there is at least one output qubit observed or operated on later. Consequently, the channels achievable via each layer of unitaries then also constitute only a subset of all possible channels between its input and output density matrices.

\section{Related Work}\label{sec:related_work}
Dephasing or decoherence was used to connect probabilistic graphical models and TNs by Miller et al.~\cite{miller_2021}. Robeva et al. showed that the data defining a discrete undirected graphical model (UGM) is equivalent to that defining a tensor network with non-negative nodes \cite{duality}. The Born machine (BM) \cite{Glasser_2019,miller_2021} is a more general probabilistic model built from TNs that arise naturally from the probabilistic interpretation of
quantum mechanics. The locally purified state (LPS) \cite{Glasser_2019} adds to the BM some purification edges each of which partially traces over a node, and represents the most general family of quantum-inspired probabilistic models. The decohered Born Machine (DBM) \cite{miller_2021} adds to a subset of the virtual bonds in BM some decoherence edges that fully dephase the underlying density matrices. A fully-DBM, i.e., a BM all of whose virtual bonds are decohered, can be viewed as a discrete UGM \cite{miller_2021}. Any DBM can be viewed as an LPS, and vice versa \cite{miller_2021}. A summary of the relative expressiveness of these families of probabilistic models is given in App.~\ref{app:pgm_tn}.

The unitary TTN and the MERA, dephased or not, are DBMs or equivalently LPSs. Each partial tracing in them is represented by a purification edge, while each dephasing channel acting on the input of a unitary node in them can be viewed as a larger unitary node contracting with some environment node and the input node, before tracing out the environment degree of freedoms using a purification edge. Each of the tensor networks produces a normalized joint probability once the data nodes are specified with normalized quantum states and the readout node is specified with a basis state. Fully-dephasing every virtual bond in the network gives rise to a fully-DBM, which can be also viewed as a discrete UGM in the dual graphical picture. We describe in App.~\ref{sec:dephasing_uni_ttn} that, by directly taking into account the effect of the partial tracing or the purification, the fully-dephased networks can also be viewed as Bayesian networks via some directed acyclic graphs (DAGs).

\section{Numerical Experiments}\label{sec:numerical_experiments}

\begin{figure*}
 \center{\includegraphics[scale=0.45]{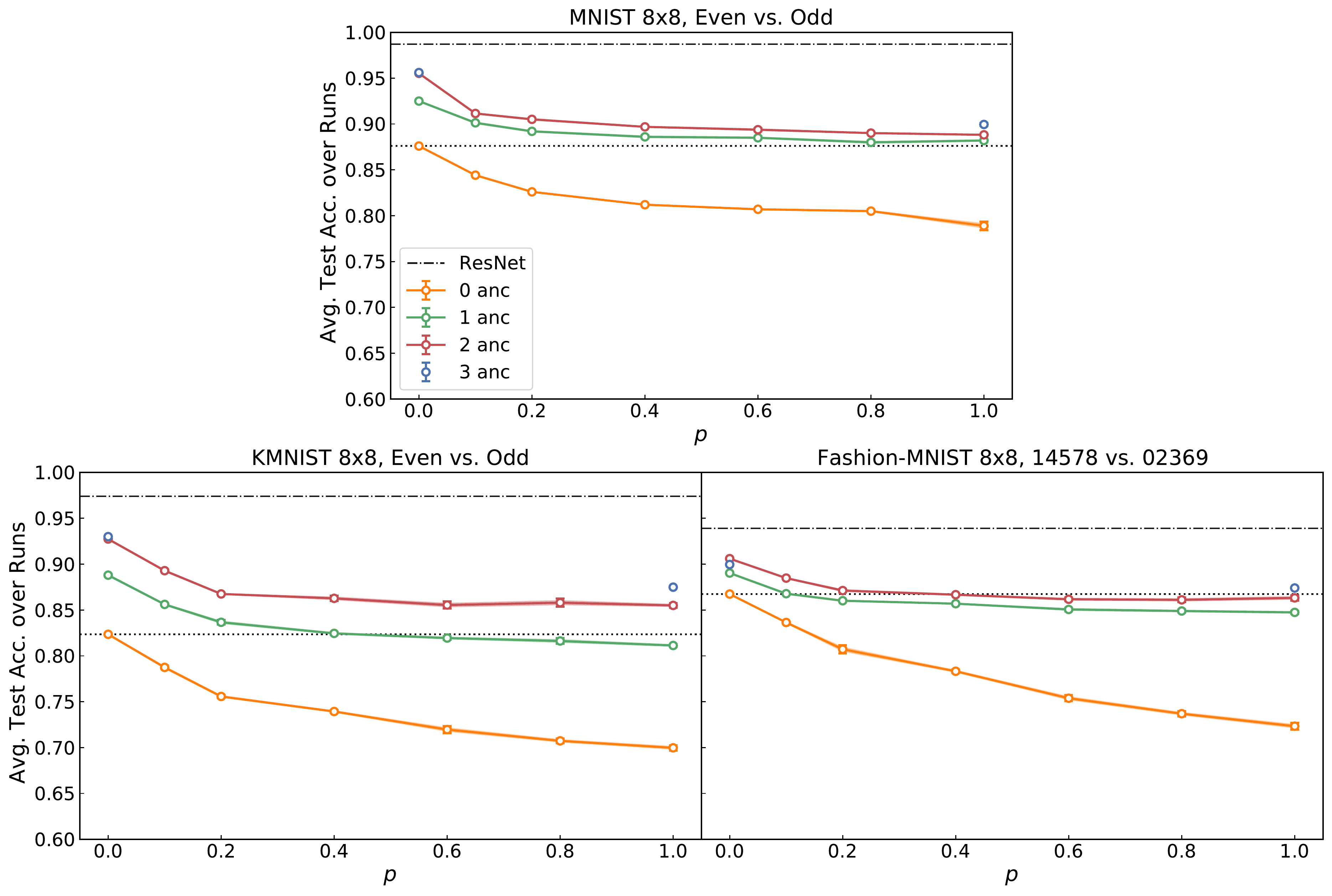}}
 \captionsetup{justification=raggedright, singlelinecheck=false}
\caption{Average testing accuracy over five runs with random batching and random initialization as a function of dephasing probability $p$ when binary-classifying $8\times8$ compressed MNIST, KMNIST, or Fashion-MNIST images. In each image dataset, we group the original ten classes into two, with the grouping shown in the titles. Every layer of the unitary TTN, including the data layer, is locally dephased with a probability $p$. Each curve represents the results from the network with a certain number of ancillas added per data qubit, with the error bars showing one standard error. The dotted reference line shows the accuracy of the non-dephased network without any ancilla.}
 \label{fig:three_datasets}
\end{figure*}

To demonstrate the competing effect between dephasing and adding ancillas while accordingly increasing the bond dimension of the network, we train the unitary TTN to perform binary classification on grouped classes on three datasets of different levels of difficulty\footnote{\url{https://github.com/HaoranLiao/dephased_ttn_mera.git}. Example images of the three datasets are shown in App.~\ref{sec:datasets}.}. 
Recall that $n_i$, $n_a$, and $n_o$ respectively denote the number of input data qubits, ancillas, and output qubits, of every unitary node in the first layer of the TN. We employ TTNs with $n_i = 2$, $n_a\in\{0, n_i, 2n_i, 3n_i\}$, and $n_o=1/2(n_i+n_a)$ for every unitary node in the first layer, and with virtual bond dimensions equal $1/2(n_i+n_a)$. We also employ MERAs with $n_i = 2$, $n_a\in\{0, n_i\}$, and $n_o=1/2(n_i+n_a)$ for every unitary node in the first layer, and with virtual bond dimensions equal $1/2(n_i+n_a)$. The root node in either network has one output qubit measured for a binary prediction.

We vary both the dephasing probability $p$ in dephasing every layer of the network, and the number of ancillas, which results in a varying bond dimension of the TTN. In the fully-dephased limit, the unitary TTN essentially becomes a Bayesian network that computes a classical joint probability distribution (see App.~\ref{sec:fully-dephase_tn}).

In each dataset, we use a training set of $50040$ samples of $8\times8$-compressed images and a validation dataset of $9960$ samples, and we employ the qubit encoding given in Eq.~\eqref{eq:qubit_encoding}. The performance is evaluated by classifying another $10000$ testing samples. The unitarity of each node is enforced by parametrizing a Hermitian matrix $H$ and letting $U=e^{iH}$. In all of our cases where the model can be efficiently simulated\footnote{If the model cannot be efficiently simulated, stochastic approximations such as the simultaneous perturbation stochastic approximation (SPSA) with momentum algorithm~\cite{Huggins_2019} can be used for training.}, they can be optimized with analytic gradients using the Adam optimizer~\cite{Kingma2015AdamAM} with respect to a categorical cross-entropy loss function, with backpropagations through the dephasing channels. Values of the hyperparameters employed in the optimizer (learning rate) and for initializion of the unitaries (standard deviations) are tabulated in App.~\ref{sec:datasets}. The ResNet-18 model~\cite{resnet}, serving as a benchmark of the state-of-the-art classical image recognition model, is adapted to and trained/tested on the same compressed, grayscale images.

For the first $8\times 8$-compressed, grayscale MNIST~\cite{mnist} dataset, and the second $8\times 8$-compressed, grayscale KMNIST~\cite{kmnist} dataset, we group all even-labeled original classes into one class and group all odd-labeled original classes into another, and perform binary classification on them. For the third $8\times 8$-compressed, grayscale Fashion-MNIST~\cite{fashion_mnist} dataset, we group $0,2,3,6,9$-labeled original classes into one class and the rest into another. The binary classification performance on each of the three datasets as a function of dephasing probability $p$ and the number of ancillas is shown for the unitary TTN in Fig.~\ref{fig:three_datasets}. Due to high computational costs, we simulate a three-ancilla network with $p$ values equal to $0$ and $1$ only. This suffices to reveal the performance trends in both the non-decohered unitary tensor network and the corresponding Bayesian network. 

\begin{figure}
 \centering
\includegraphics[scale=0.4]{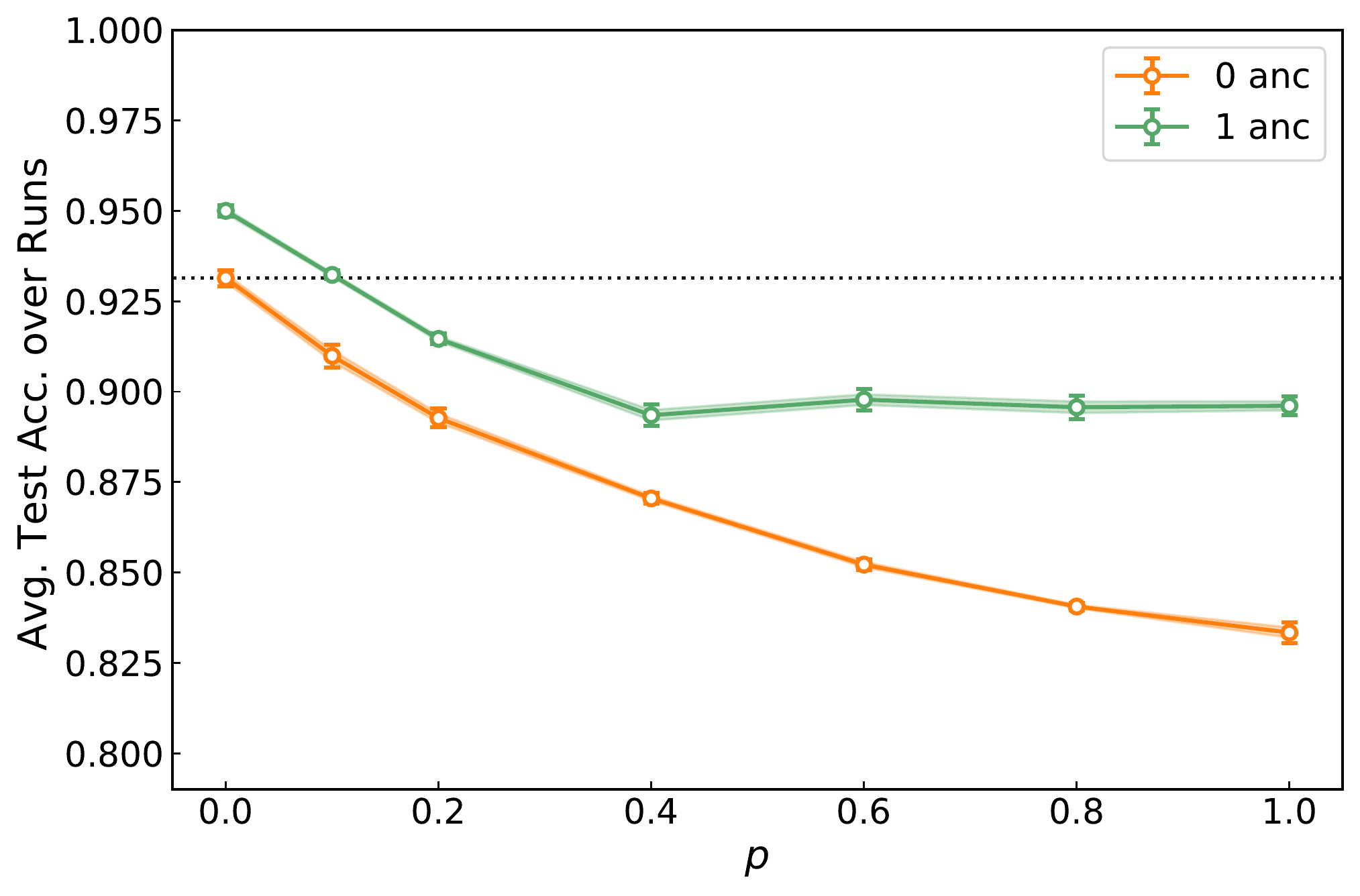}
\captionsetup{justification=raggedright}
 \caption{Average testing accuracy over ten runs with random batching and initialization as a function of dephasing probability $p$ in dephasing a 1D MERA structured tensor network to classify the eight principle components of non-compressed MNIST images. Ancillas are added per data qubit. The dotted reference line shows the accuracy of the non-dephased network without any ancilla.}
 \label{fig:mera_mnist8pca}
\end{figure}

There are two interesting observations to make on the results in Fig.~\ref{fig:three_datasets}. First, the classification performance is very sensitive to small decoherence and decreases the most rapidly in the small $p$ regime, especially in networks with at least one ancilla added. Further dephasing the network does not decrease the performance significantly, and in some cases, it does not further decrease the performance at all. A similar observation is made for the MERA (see Fig.~\ref{fig:mera_mnist8pca}). Second, in the strongly dephased regime where the ancillas are very noisy, adding such noisy ancillas helps the network regain performance relative to that of the non-dephased no-ancilla network. On all three datasets, the performance regained after adding two ancillas across all dephasing probabilities is comparable to the performance with the no-ancilla non-dephased network. This suggests that in implementing such unitary TTNs in the NISQ era with noisy ancillas, it is favorable to add at least two ancillas to the network and to accordingly expand the bond dimension of the unitary TTN to at least eight, regardless of the decoherence this may introduce.

However, due to the high computational costs with more than three ancillas added to the network, our experiments do not provide sufficient information about whether the corresponding Bayesian network in the fully-dephased limit will ever reach the same level of classification performance as the non-dephased unitary TTN by increasing the number of ancillas. Despite this, we note that in the KMNIST and Fashion-MNIST datasets the rate of improvement of the Bayesian network as more ancillas are added is diminishing.

Fig.~\ref{fig:three_datasets} shows that when classifying the Fashion-MNIST dataset, adding three ancillas in the non-decohered network leads to a slightly worse performance than just adding two ancillas. This may be attributed to the degradation problem in optimizing complex models, which is well-known in the context of classical neural networks~\cite{resnet}. For neural networks, this is manifested by a performance drop in both training and testing as more layers are added, and is distinguished from overfitting where only the testing accuracy drops. In the current unitary TTN calculations, the eight-qubit unitaries that arise in the three-ancilla setting are significantly harder to optimize than the six-qubit unitaries that arise in the two-qubit setting. The optimization was unable to adequately learn the eight-qubit unitaries and thus there is a small drop in performance seen on increasing the ancilla count from two to three.

\begin{figure}
 \centering
\includegraphics[scale=0.41]{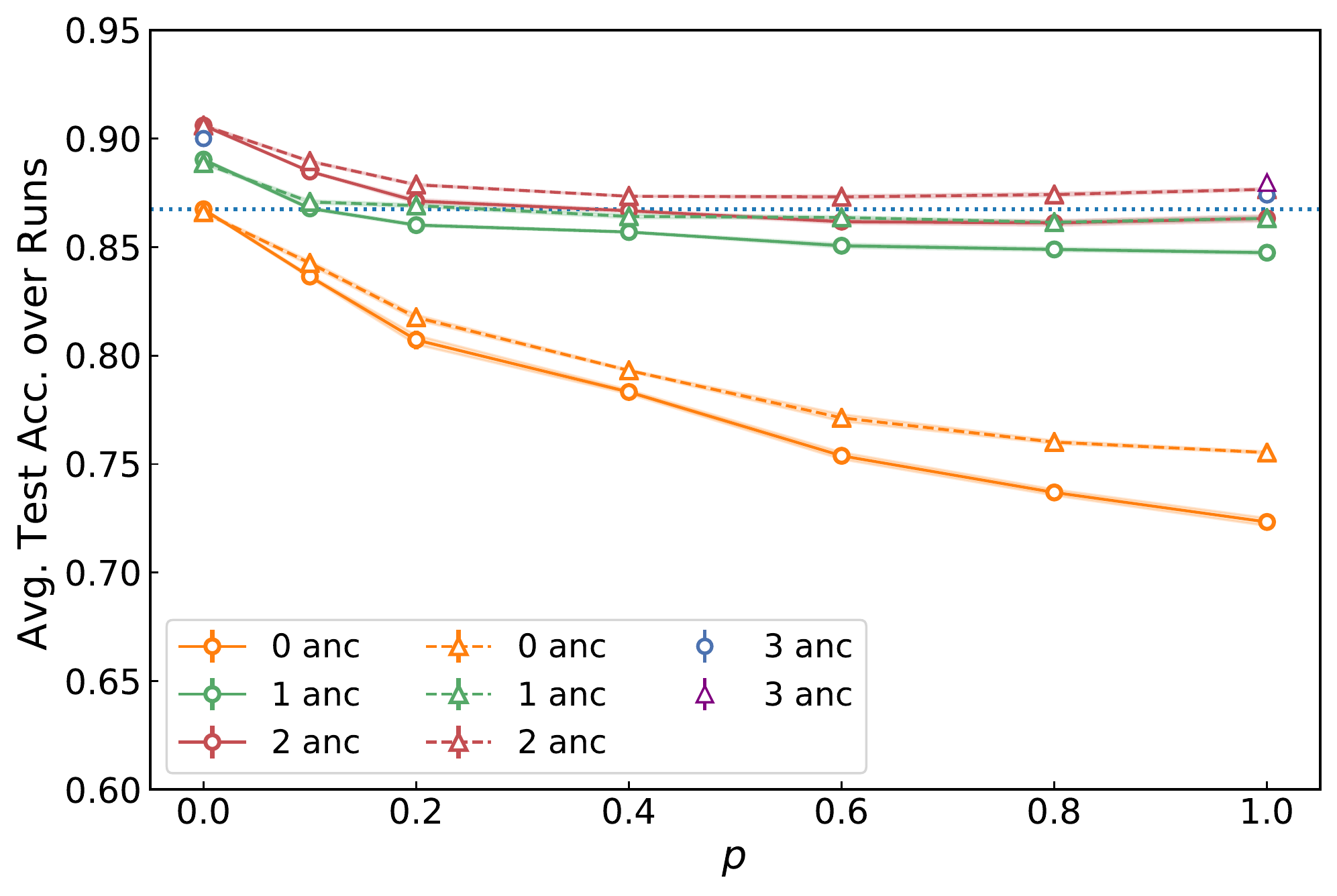}
\captionsetup{justification=raggedright}
 \caption{Average testing accuracy over five runs as a function of dephasing probability $p$ when classifying $8\times8$ compressed Fashion-MNIST images. Each curve represents the results from the network with a certain number of ancillas added per data qubit. The circles (triangles) show the performance of the unitary TTN when every layer including (except) the data layer is locally dephased with a probability $p$. The dotted reference line shows the accuracy of the non-dephased network without any ancillas.}
 \label{fig:deph_net_only}
\end{figure}

Dephasing the data layer is special compared to dephasing other internal layers within the network, since the coherence in each of the product-state data qubits has not been mixed to form the next-layer features. Since the coherences are non-linear functions of the diagonals of $\rho$, given the linear nature of tensor networks, it is not possible to reproduce the coherence in the data qubits in subsequent layers once the input qubits are fully-dephased. To examine to what extent the observed performance decrement may be attributed to decoherence within the network as opposed to decoherence of the data qubits, we perform the same numerical experiment on the Fashion-MNIST dataset but keep the input qubits coherent without any dephasing. The result, shown in Fig.~\ref{fig:deph_net_only}, indicates that the decoherence of the virtual bonds in the unitary TTN alone is a significant source causing the classification performance to decrease, accounting for more than half of the performance decrement.

\section{Discussion}\label{sec:discussion}
In this paper, we investigated the competition between dephasing tensor network QML models and adding ancillas to the networks, in an effort to investigate the advantage of coherence in QML and to provide guidance in determining the number of noisy ancillas to be included in NISQ-era implementations of these models. On one hand, as we increase the dephasing probability $p$ of every layer of the network, every regressor associated with each layer of unitary nodes will have certain terms in it damped by some power of $(1-p)$. The damping cannot be offset by the regression coefficients which are given in terms of the elements of the unitary matrices. The effect of this damping of the regressors under dephasing decreases the classification accuracy of the QML model. When the network is fully-dephased, these regressors are eliminated, and the tensor network QML model becomes a classical Bayesian network that is completely describable by classical probabilities and stochastic matrices. On the other hand, as we increase the number of input ancillas and accordingly increase the virtual bond dimensions of the tensor network, we allow the network to represent a larger subset of unitaries between the input and output qubits. As a result, the performance of the network improves, as demonstrated by adding up to two ancillas and a corresponding increment of the virtual bond dimension to eight in our numerical experiments. This improvement applies to all decoherence probabilities. We also find that adding more than two ancillas gives either diminishing or no improvement (Fig.~\ref{fig:three_datasets}). The numerical experiments are insufficient to show whether the performance of the corresponding Bayesian network can match that of the non-decohered network as more than three ancillas are added, although we did find that in the KMNIST and Fashion-MNIST datasets the rate of improvement of the Bayesian network as more ancillas are added is diminishing. It remains an open question where coherence provides any quantum advantage in QML. 

Most importantly, we find that the performance of the two-ancilla Bayesian network, namely the fully-dephased network, is comparable to that of the corresponding non-decohered unitary TTN with no ancilla, suggesting that when implementing the unitary TTN, it is favorable to add at least two arbitrarily noisy ancillas and to accordingly increase the virtual bond dimension to at least eight. 

We also observe that the performance of both the unitary TTN and the MERA decreases most rapidly in the small decoherence regime. With ancillas added, the performance decreases and quickly levels off at around $p=0.2$ for the unitary TTN. The MERA with one ancilla added also exhibits this level-off performance after around $p=0.4$. However, without any ancilla added, neither the unitary TTN nor the MERA shows a level-off performance and their performance decreases all the way until the networks are fully dephased. This contrast is an interesting phenomenon to be studied in the future. 

We note that the ancilla scheme discussed in Sec.~\ref{sec:ancillas} and the theoretical analysis of the fully-decohered network presented in App.~\ref{sec:fully-dephase_tn} are also relevant to other variational quantum ansatz states beyond tensor network QML models. For example, the analysis applies to non-linear QML models consisting of generic unitaries, such as those incorporating operations conditioned on mid-circuit measurement results of some of the qubits~\cite{Cong2019}. They may behave similarly under the competition between decoherence and adding ancillas, and it is an interesting problem for future investigation.

\section*{Acknowledgments}
We would like to thank William J. Huggins for his insights into the problem and helpful discussions. We also thank the Google Cloud Research Credits program for providing cloud hardware for our numerical experiments.

\section*{Declarations}
\subsection{Competing Interests}
The authors have no conflicts of interest to declare that are relevant to the content of this article.

\subsection{Authors' contributions}
H.L. wrote the manuscript text, prepared the figures, and contributed to most of the numerical experiments and part of the theoretical analysis. I.C. contributed to part of the theoretical analysis and numerical experiments. Z.Y contributed to part of the numerical experiments. K.B.W. contributed to part of the theoretical analysis and to the writing of the manuscript. All authors have reviewed the manuscript.

\subsection{Funding}
This material is based upon work supported by the UC Noyce Initiative and the U.S. Department of Energy, Office of Science, National Quantum Information Science Research Centers, Quantum Systems Accelerator. 

\subsection{Availability of data and materials}
The source code of both the unitary TTN and the MERA discriminative QML models, as well as the datasets for the numerical experiments, are available at \url{https://github.com/HaoranLiao/dephased_ttn_mera.git}.

\bibliography{refer}

\onecolumngrid
\section*{Appendices}
\begin{appendix}

\section{Discrete Bayesian Networks}\label{sec:bayes_net}
Let a set of vertices and an edge set of ordered pairs of vertices form a directed graph $G=(V,E)$, and let $X=\{X_v\}, \forall v\in V$ be a set of discrete random variables indexed by the vertices. Let $\text{pa}(v)$ or $X_{\text{pa}(v)}$ denote the set of parent vertices/variables each of which has an edge directed towards $v$. A directed edge represents some conditional probability of the variable on its parent. We say that $X$ is a discrete Bayesian network (a.k.a. belief network) with respect to $G$ if $G$ is acyclic, namely, it is a directed acyclic graph (DAG), or equivalently if the joint probability mass function of $X$ can be written as a product of the individual probability mass functions conditioned on their parent variables, i.e., $P(X)=\prod_{v\in V}P(X_v|X_{\text{pa}(v)})$.

\section{Fully-dephased Unitary Tensor Networks}\label{sec:fully-dephase_tn}
\subsection{Fully-dephasing Qubits after Unitary Evolution}
 To fully dephase a quantum state, we simply choose a basis to represent the density matrix and then set all off-diagonal elements of the matrix to zero, leaving the diagonal elements unchanged. If we represent the fully-dephasing ($p=1$) superoperator as $\mathcal{D}$, then
\begin{equation}\label{eq:fully_dep}
 \mathcal{D}[\rho] = \sum_i \braket{i|\rho|i}\ketbra{i}{i} = \sum_i \rho_{ii}\ketbra{i}{i}.
\end{equation}
For convenience, we adopt the notation $\lambda_i \equiv \rho_{ii}$, where the $\lambda_i$ can be identified as probabilities from some discrete distribution. If we allow a generic unitary $U$ to act on $\rho$ before it is fully-dephased, then we have
\begin{equation}
\begin{split}
    \mathcal{D}[U\rho U^{\dagger}] &= \sum_i \braket{i|U\rho U^{\dagger}|i}\ketbra{i}{i} = \sum_{ijk} \rho_{jk}\braket{i|U|j}\braket{k|U^{\dagger}|i}\ketbra{i}{i},
\end{split}
\end{equation}
so that the new probabilities $\lambda'_i$ encoded in the fully-dephased state are given by
\begin{equation}\label{eq:fully_dep_after_U}
\begin{split}
    \lambda'_i &= \mathcal{D}[U\rho U^{\dagger}]_{ii} =\sum_{jk}\rho_{jk}\braket{i|U|j}\braket{k|U^{\dagger}|i}=\sum_{jk}\rho_{jk}U_{ij}U^*_{ik}
\end{split}
\end{equation}
From Eq.~\eqref{eq:fully_dep_after_U}, we can see that each probability is a function of the entire density matrix, along with the elements of $U$. If $\rho$ is assumed to be fully-dephased already, then $\rho_{jk} = \lambda_{j}\delta_{jk}$ and therefore
\begin{equation}
    \lambda'_i = \sum_{jk}\lambda_j\delta_{jk}U_{ij}U^*_{ik} = \sum_j \lambda_j\abs{U_{ij}}^2 = \sum_j M_{ij}\lambda_j.
\end{equation}
By the unitarity of $U$, $M_{ij} \equiv |U_{ij}|^2$ is doubly stochastic, i.e., $\sum_i M_{ij}=\sum_i\abs{U_{ij}}^2 = \mathbbm{1}_j$ and $\sum_j M_{ij}=\sum_j\abs{U_{ij}}^2 = \mathbbm{1}_i$, which maps the old probabilities $\lambda$ to new probabilities $\lambda'$ that are normalized, i.e., $\sum_i \lambda'_i=\sum_{ij}M_{ij}\lambda_j=\sum_j\mathbbm{1}_j\lambda_j=1$. Such doubly stochastic matrices $M$ that correspond to some unitaries are called unitary-stochastic matrices. For $N\leq 2$, all $N\times N$ doubly stochastic matrices are also unitary-stochastic. But unitary-stochastic matrices form a proper subset of doubly stochastic matrices for $N\geq 3$\footnote{The dimension of the parameter space for $N\times N$ unitary-stochastic matrices is $(N-1)^2$ as for doubly stochastic matrices. The parameter space covered by unitary-stochastic matrices is, however, in general, smaller than that covered by doubly stochastic matrices \cite{Tanner_2001}.} \cite{Zyczkowski_2003, Tanner_2001}.

\subsection{Fully-dephasing a Reduced Density Matrix after Unitary Evolution}\label{sec:dephase_reduced_rho}

In some tensor networks such as the TTN, the effective size of the feature space is reduced by tracing over some of the degrees of freedom after each layer. The combined effects of the unitary layer and partial trace produce a quantum channel, whose output is then fully-dephased. If we partition the Hilbert space of an input density matrix $\rho$ into parts $A$ and $B$, then the outputs $\lambda'_{i_B}$ after tracing over part $A$ are given by
\begin{equation}
\begin{split}
    \lambda'_{i_B} & =
    \left[\Tr_A\left(\mathcal{D}[U\rho U^{\dagger}]\right)\right]_{i_Bi_B} \\
    &= \left[\sum_{i_Ai_Bjk}\Tr_A\left(\rho_{jk}\braket{i_A i_B|U|j}\braket{k|U^{\dagger}|i_A i_B}\ketbra{i_A}{i_A}\ketbra{i_B}{i_B}\right)\right]_{i_Bi_B}
    \\
    &=\sum_{i_Ajk}\rho_{jk}\braket{i_A i_B|U|j}\braket{k|U^{\dagger}|i_A i_B}\Tr\left(\ketbra{i_A}{i_A}\right)\\
    &=\sum_{jk}\rho_{jk}\sum_{i_A} U_{i_Ai_Bj}U^*_{i_Ai_Bk}.
\end{split}
\end{equation}
We can again see that the output diagonals depend on all elements of $\rho$ and $U$. If $\rho$ is already fully dephased, then we have
\begin{equation}
\begin{split}
    \lambda'_{i_B} &= \sum_{jk}\lambda_j\delta_{jk}\sum_{i_A} U_{i_Ai_Bj}U^*_{i_Ai_Bk} 
    = \sum_j \lambda_j \sum_{i_A}\abs{U_{i_Ai_Bj}}^2 = \sum_j S_{i_Bj}\lambda_j,
\end{split}
\end{equation}
where $S_{i_Bj} \equiv \sum_{i_A}\abs{U_{i_Ai_Bj}}^2$ is a rectangular singly stochastic matrix with respect to index $i_B$ only, i.e., $\sum_{i_B} S_{i_Bj}=\sum_{i_Ai_B}|U_{i_Ai_Bj}|^2=\mathbbm{1}_j$. It again maps the old probabilities $\lambda$ to new probabilities $\lambda'$ which are normalized, i.e., $\sum_{i_B} \lambda'_{i_B}=\sum_{i_Bj}S_{i_Bj}\lambda_{j}=\sum_j \mathbbm{1}_j\lambda_j=1$. We remark that the output index $i_B$ runs from $1$ to $\dim(B)$, while the input index $j$ runs from $1$ to $\dim(A)\cdot\dim(B)$, and the Bayesian update by this singly stochastic matrix applies only in the coarse-graining direction.

\subsection{Fully-dephasing the Unitary TTN}\label{sec:dephasing_uni_ttn}
Dephasing a unitary TTN is to apply local dephasing channels on each pair of output bonds before contracting with the node at the next layer, as shown in Fig.~\ref{fig:ttn} (right). In terms of the underlying density matrix, the dephasing channel is to apply Eq.~\eqref{eq:multi_qubits_dephase} to the bonds, each of which may represent a higher-dimensional state if there are ancilla qubits added as discussed in Sec.~\ref{sec:ancillas}. We note that assuming local dephasing, there is no need to dephase before partially tracing out some generally entangled qubits out of the unitary TTN node, say tracing over part $A$ of the output system $AB$, since there exists a $U_{AE}$ on $\rho_{AB}\otimes\rho_E$ by the definition of dephasing such that
\begin{equation}
\begin{split}
    \Tr_A\left(\mathcal{E}_A\left[\rho_{AB}\right]\right)&=\Tr_A\left[\Tr_E\left(U_{AE}\rho_{AB}\otimes\rho_EU_{AE}^\dagger\right)\right]
    = \Tr_A(\rho_{AB}).
\end{split}
\end{equation}
A diagram of the dephased unitary TTN is shown in Fig.~\ref{fig:ttn} (right).

\begin{figure}[H]
 \centering
\includegraphics[scale=0.23]{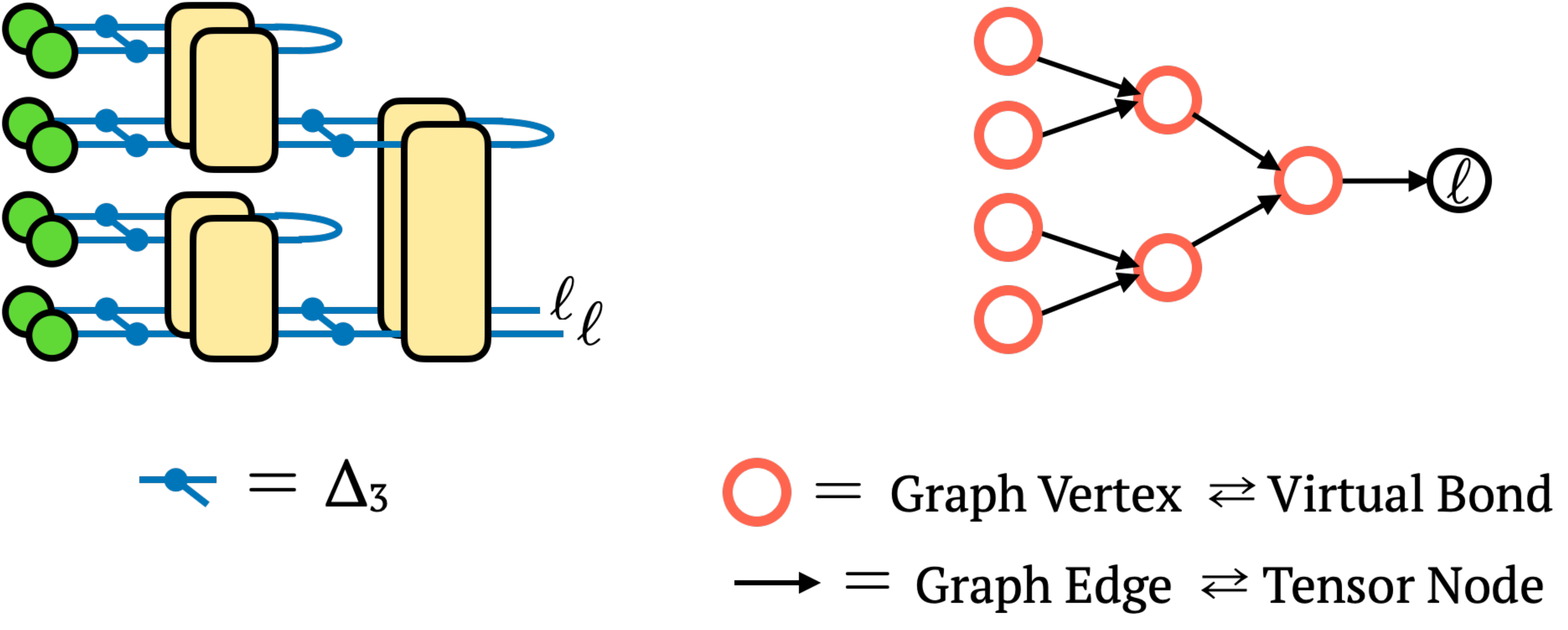}
\captionsetup{justification=raggedright, singlelinecheck=false}
 \caption{Left: Fully-dephasing a unitary TTN, where the third-order copy tensor $\Delta_3$ is defined as $\Delta_3=\sum_i e_i^{\otimes 3}$ with $e_i$ the qubit basis state (see App.~\ref{sec:copy_tensor}). Right: The dual graphical picture of the fully-dephased unitary TTN as a Bayesian network via a directed acyclic graph (DAG). The transition matrices conditioning on each pair of input vectors are rectangular singly stochastic matrices $S$'s reduced from some unitary-stochastic matrices.}
 \label{fig:dephased_ttn}
\end{figure}

 As shown in App.~\ref{sec:dephase_reduced_rho}, fully decohering after partially tracing out every composite node of a unitary TTN leads to a TTN composed of nodes each of which is a rectangular singly stochastic matrix $S$ (reduced from a unitary-stochastic matrix), acting on a vector of the diagonals of a density matrix, that only preserves the normalization in the coarsed-graining direction. The fully-dephased TTN then exhibits a chain of conditional probabilities and can be interpreted as successive Bayesian updates across layers. A diagram using the third-order copy tensors (see App.~\ref{sec:copy_tensor}) to fully dephase the unitary TTN is shown in Fig.~\ref{fig:dephased_ttn} (left), and the dual graphical picture as a Bayesian network is depicted in Fig.~\ref{fig:dephased_ttn} (right).

 Formally, a fully-dephased unitary TTN can be viewed as a discrete Bayesian network via a DAG with input quantum states as parent variables. In other words, the Bayesian network provides a dual graphical formulation of the fully-dephased unitary TTN, with the graph edges functioning as the tensor nodes while the graph vertices acting as the virtual bonds \cite{duality, miller_2021}. The graph vertices in the Bayesian network, which is dual to the virtual bonds in the TTN composed of stochastic matrices, represent vector variables $\lambda^{(k,j)}\equiv\text{diag}(\rho^{(k,j)})$, where $k$ and $j$ denotes the $j$-indexed vertices at the $k$th layer of the network with $0$ indexing the layer with parent variables, and $\rho$ is the corresponding density matrix in the dual tensor network picture. We use the shorthand $\lambda^{(k)}\equiv\{\lambda^{(k,0)},\dots,\lambda^{(k,n_k)}\}$ to group all $n_k$ vertices at the $k$th layer into a set. The output vertex of the Bayesian network stands for a readout variable $\ell$ specifying the basis state of the measurement. The Bayesian network then yields a joint probability once the parent variables are specified with normalized quantum states, i.e., the joint probability represented by the network can be written in the following factorized form
\begin{equation}\label{eq:bayesian_network_factorization}
\begin{split}
    P&(\lambda^{(0)}\dots,\lambda^{(\log(m))}, \ell)
    =P(\ell|\lambda^{(\log(m))})\prod_{k=1}^{\log(m)} P(\lambda^{(k)}|\lambda^{(k-1)})P(\lambda^{(0)}),
\end{split}
\end{equation}
where $m\equiv n_0$ is the number of vertices at the data layer. $P(\lambda^{(k)}|\lambda^{(k-1)})$ is the conditional probability represented by the edges between the ($k-1$)th and $k$th layer of the Bayesian network, or equivalently by the rectangular singly stochastic matrices at the $k$th layer of the dual tensor network. $P(\ell|\lambda^{(\log(m))})$ is the conditional probability of obtaining the basis vector $\ell$.

When, for instance, the unitary TTN is fully dephased to become a Bayesian network, both schemes of adding ancillas, as described in Sec.~\ref{sec:ancillas}, give rise to networks that share the same form of factorized conditional probabilities as shown in Eq.~\eqref{eq:bayesian_network_factorization}. The difference between the two schemes lies in that adding ancillas per node leads to $\lambda^{k, j}$ fixed at two dimensional $\forall k, j$, whereas adding ancillas per data qubit allows $\lambda^{k, j}$'s dimension to grow with the number of ancillas $\forall k\in\{1,\dots,\log(m)\}, \forall j$, since increasing virtual bond dimension increases the number of diagonals.

\subsection{Fully-dephasing the MERA} \label{sec:dephasing_mera}
\begin{figure}
 \centering
\includegraphics[scale=0.30]{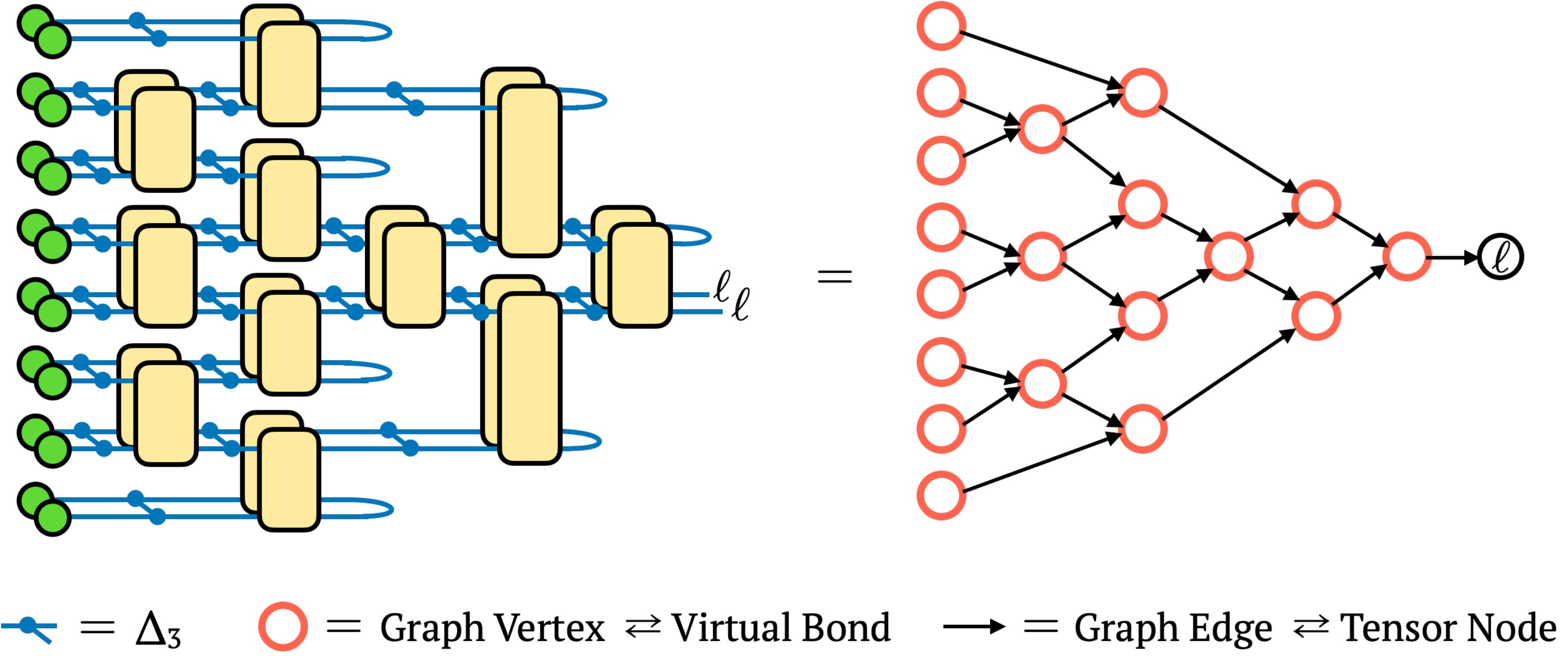}
\captionsetup{justification=raggedright, singlelinecheck=false}
 \caption{Left: Fully-dephasing a MERA. Right: Equivalently, the dual graphical picture of the fully-dephased unitary TTN as a Bayesian network via a DAG, since the fully-dephased MERA is a tensor network composed of unitary-stochastic matrices $M$'s and rectangular singly stochastic matrices $S$'s with respect to the coarse-graining direction, with input being the diagonals of the encoded qubits.}
 \label{fig:dephased_mera}
\end{figure}

Similar to the fully-dephased unitary TTN, the fully-dephased MERA is shown in Fig.~\ref{fig:dephased_mera} (left), whose dual graphical formulation as a Bayesian network is shown in Fig.~\ref{fig:dephased_mera} (right), such that the joint probability yielded by the network upon specifying the input quantum states as the parent variables has the same factorized form as Eq.~\eqref{eq:bayesian_network_factorization}. An entangler with fully-dephased input and output transforms to a unitary-stochastic matrix $M$, and the partially-traced-over unitary, serving as the ``isometry'', with fully-dephased input and output transforms to a singly stochastic matrix $S$ (reduced from a unitary-stochastic matrix) with respect to the coarse-graining direction. We note that the dimension of the vector variables dual to the output bonds of entanlgers in the tensor network picture is twice as large as other variables, since they represent correlated variables outputted by the unitary-stochastic matrices. Each of the two outgoing directed edges from these variables can be interpreted as a conditional probability conditioning on half of the support of these discrete variables.

\section{Ancillas Are Required to Achieve Evolution by Singly Stochastic Matrices}\label{sec:stinespring}
Ancillas are necessary to allow the evolution of fully-dephased input induced by a generic unitary to be as expressive as that induced by general singly stochastic matrices. Consider a singly stochastic matrix
\begin{equation}\label{eq:stochastic_matrix}
\begin{bmatrix}
    1 & 1 & 1 & 1\\
    0 & 0 & 0 & 0\\
  \end{bmatrix},
\end{equation}
which maps an input state in $\{\ket{00}, \ket{01}, \ket{10}, \ket{11}\}$ to $\ket{0}$. Note that this is naturally a mapping between fully-dephased input and fully-dephased output. But this mapping cannot be achieved by acting a unitary on the data qubit alone. To achieve that, we need to unitarily evolve a combined system including at least one ancilla. After tracing out the ancilla, it is possible to leave the data qubit in $\ket{0}$. Namely, $\{\ket{00}, \ket{01}, \ket{10}, \ket{11}\}\rightarrow \ket{0}\otimes\ket{0}_E$ or $\{\ket{00}, \ket{01}, \ket{10}, \ket{11}\}\rightarrow \ket{0}\otimes\ket{1}_E$ is achievable by a unitary on the combined system. Note that this is also a mapping between fully-dephased input and fully-dephased output naturally. Therefore, considering generic unitary evolution such as contracting with a node in the unitary TTN, it is necessary to include ancillas to achieve what can be mapped by a singly stochastic matrix between the fully-dephased input and fully-dephased output.

\section{Probabilistic Graphical Models and Tensor Networks}\label{app:pgm_tn}
It was shown by Robeva et al.~\cite{duality} in Theorem 2.1 that the data defining a discrete undirected graphical model (UGM) is equivalent to that defining a tensor network (TN) with non-negative nodes, but with dual graphical notations that interchange the roles of nodes and edges. Hence, we have discrete UGM$=$non-negative TN, where $=$ represents that the two classes of model can produce the same probability distribution using the same number of parameters, i.e., they are equally expressive.

The Born machine (BM) \cite{Glasser_2019,miller_2021}, which models a probability mass function as the absolute value squared of a complex function, is a family of more general probabilistic models built from TNs that arise naturally from the probabilistic interpretation of
quantum mechanics. The locally purified state (LPS), first discussed by Glasser et al.~\cite{Glasser_2019} and generalized by Miller et al.~\cite{miller_2021}, adds to each node in a BM a purification edge, allowing it to represent the most general family of quantum-inspired probabilistic models. Glasser et al.~\cite{Glasser_2019} showed that LPS is more expressive than BM, i.e., LPS$>$BM.

The decohered Born Machine (DBM) was introduced by Miller et al.~\cite{miller_2021}, which adds to a subset of the virtual bonds BM decoherence edges that fully dephase the underlying density matrices. A BM all of whose virtual bonds are decohered is called a fully-DBM. Miller et al.~\cite{miller_2021} showed that fully decohering a BM gives rise to a discrete UGM, and conversely any subgraph of a discrete UGM can be viewed as the fully-decohered version of some BM. Hence, we have fully-DBM$=$discrete UGM.

Theorem 3 and 4 by Miller et al.~\cite{miller_2021} showed that LPS$=$DBM, since each purification edge joining a pair of LPS cores can be expressed as a larger network of copy tensors, and each decoherence edge of a DBM can be absorbed into nearby pair of tensors and form a purification edge. Following this view of LPS$=$DBM and the fact that LPS$>$BM, one arrives at DBM$>$BM, which can also be understood as BM being a special case of DBM with an empty set of decohered edges added.

A summary of the relative expressiveness is given in Tab.~\ref{tab:relative_expressiveness}.
\begin{table}[H]
\caption{The relative expressiveness, defined as the probability distributions a model can produce with the same number of parameters, among the discrete graphical model (UGM), the tensor network (TN) with non-negative nodes, the Born machine (BM), the decohered Born machine (DBM), and the locally purified state (LPS).}
\vspace{5pt}
\centering
\begin{tabular}{c|c}
    Relative Expressiveness & Ref. \\
    \hline
    discrete UGM $=$ non-negative TN & \cite{duality}\\
    fully-DBM $=$ discrete UGM & \cite{miller_2021}\\
    LPS $>$ BM & \cite{Glasser_2019}\\
    LPS $=$ DBM $>$ BM & \cite{miller_2021}\\
\end{tabular}
\label{tab:relative_expressiveness}
\end{table}

\section{Copy Tensors}\label{sec:copy_tensor}
\begin{figure}[H]
 \centering
\includegraphics[scale=0.27]{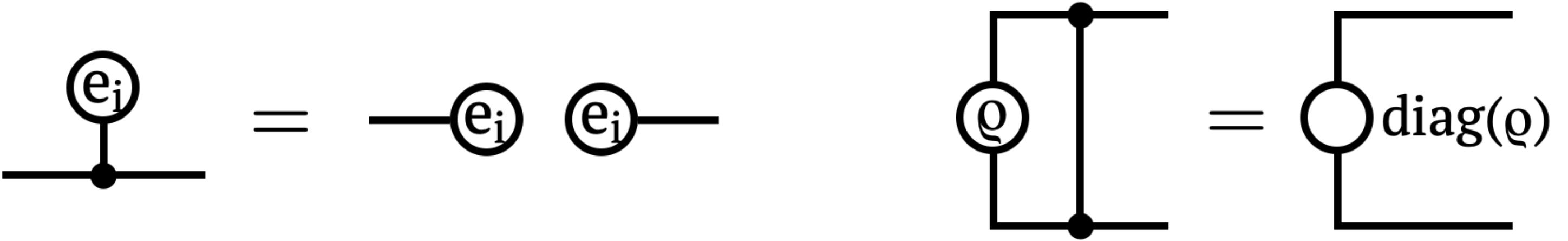}
 \caption{Left: using a third-order copy tensor contracting with a basis state vector results in an outer product of the basis vector, which can be thought of as conditioning on the same basis state upon contraction with two nodes. Right: Obtaining the diagonals of a density matrix, or a matrix in general, can be done by contracting the matrix with two third-order copy tensors and contracting one bond of each of the copy tensors together. }
 \label{fig:copy_tensor}
\end{figure}

A copy tensor of order $n$ is defined to be $\Delta_n=\sum_i e_i^{\otimes n}$ where $e_i$ is the $i$th basis vector, whose conventional tensor diagram is given as a solid dot with $n$ bonds \cite{nutshell}. An order-one copy tensor contraction can be viewed as a marginalization, while an order-three copy tensor can be used to denote conditioning on the same vector, as shown in Fig.~\ref{fig:copy_tensor}. The contraction of two third-order copy tensors with a density matrix and with themselves while leaving two bonds uncontracted conveniently reproduces Eq.~\eqref{eq:fully_dep}, in which the basis vector is the basis state $\ket{i}$, as taking the diagonals of a matrix. Therefore, it is useful to denote a dephasing channel with a dephasing rate $p=1$, as shown in Fig.~\ref{fig:copy_tensor}.

\section{Comparing the Two Ancilla Schemes in the Unitary TTN}\label{app:adding_ancillas}
As shown in Tab.~\ref{tab:ancilla_scheme_compare}, adding one ancilla per data qubit and accordingly doubling the virtual bond dimension yields superior performance to adding two ancillas per unitary node, in the task of classifying $1902$ $8\times8$-compressed MNIST images each showing a digit $3$ or $5$. Both ancilla-added unitary TTNs are trained on $5000$ samples using the Adam optimizer and validated on $2000$ samples. The two ancilla schemes share the same number of trainable parameters.

\begin{table}[H]
\caption{Average testing accuracies over five trials between adding two ancillas per unitary node and adding one ancilla per data qubit, when the dephasing rate $p=0$ or $p=1$, in the same classification task. }
\vspace{5pt}
\centering
\begin{tabular}{c|c|c}
        & Per unitary node & Per data qubit \\
    \hline
    $p=0$ & $0.938\pm0.001$ & $0.972\pm0.001$\\
    $p=1$ & $0.912\pm0.002$ & $0.940\pm0.002$\\
\end{tabular}
\label{tab:ancilla_scheme_compare}
\end{table}

\section{Datasets and Hyperparameters for the Numerical Experiments}\label{sec:datasets}
\begin{figure}
  \centering
  \subfloat[$8\times 8$-compressed MNIST images]{\includegraphics[scale=0.25]{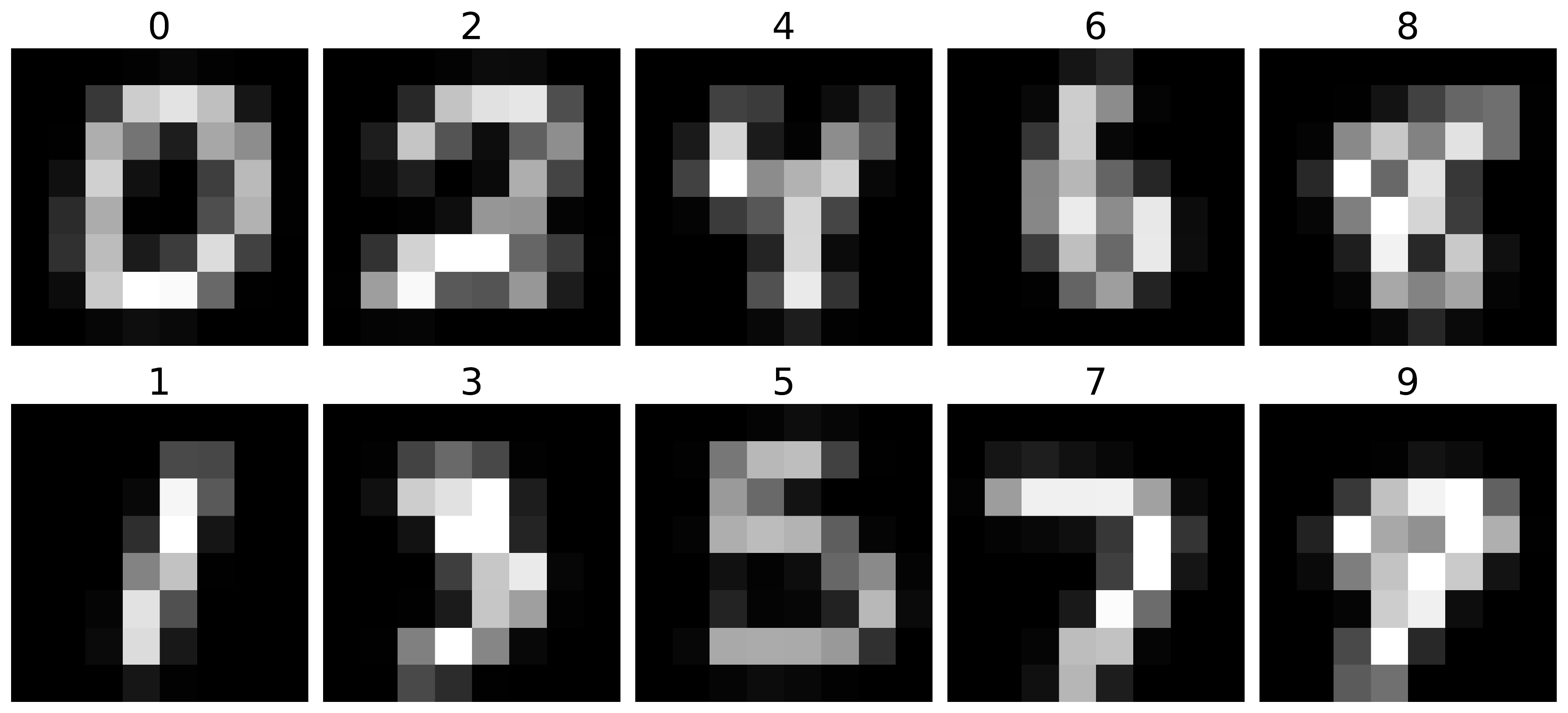}\label{fig:1}}

  \subfloat[$8\times 8$-compressed KMNIST images]{\includegraphics[scale=0.24]{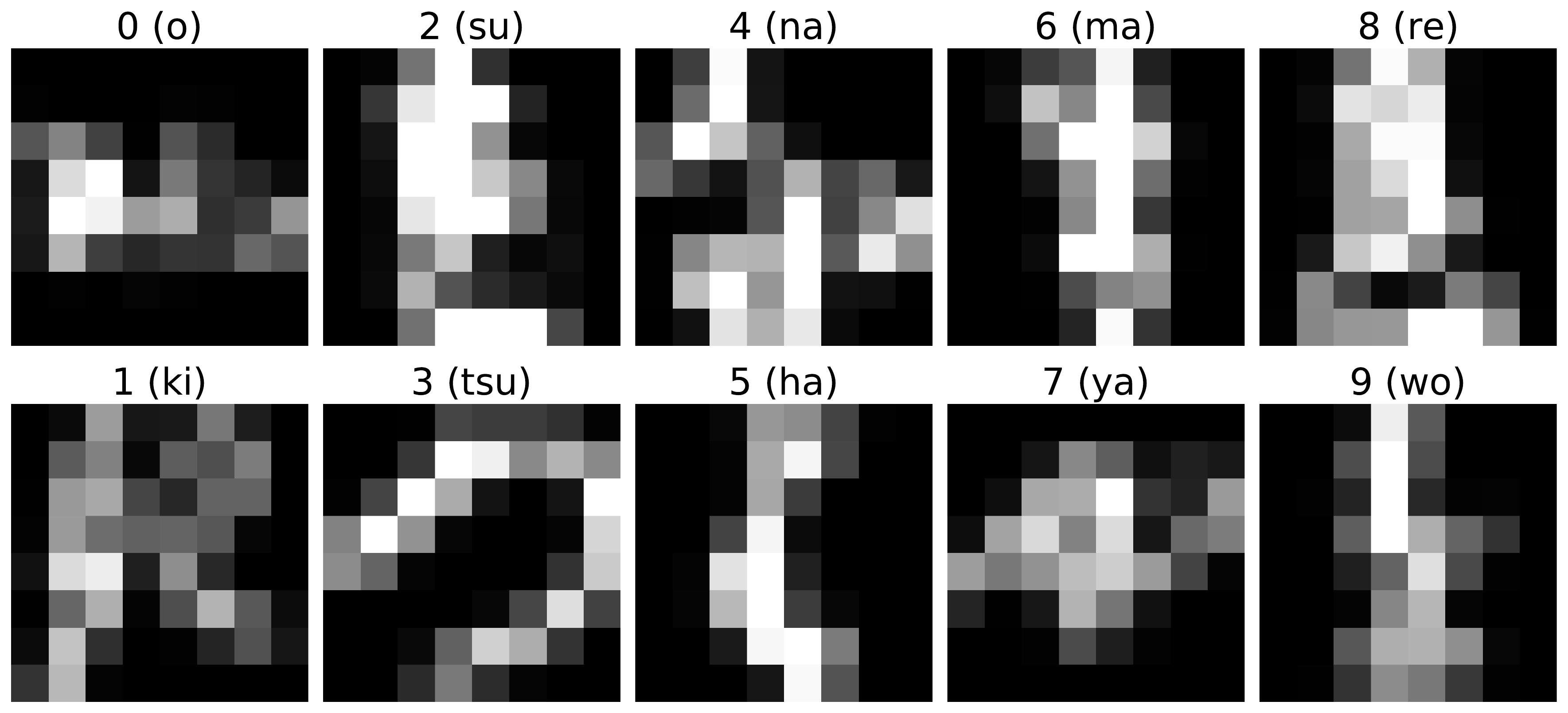}\label{fig:2}}\hspace{1em}
  \subfloat[$8\times 8$-compressed Fashion-MNIST images]{\includegraphics[scale=0.24]{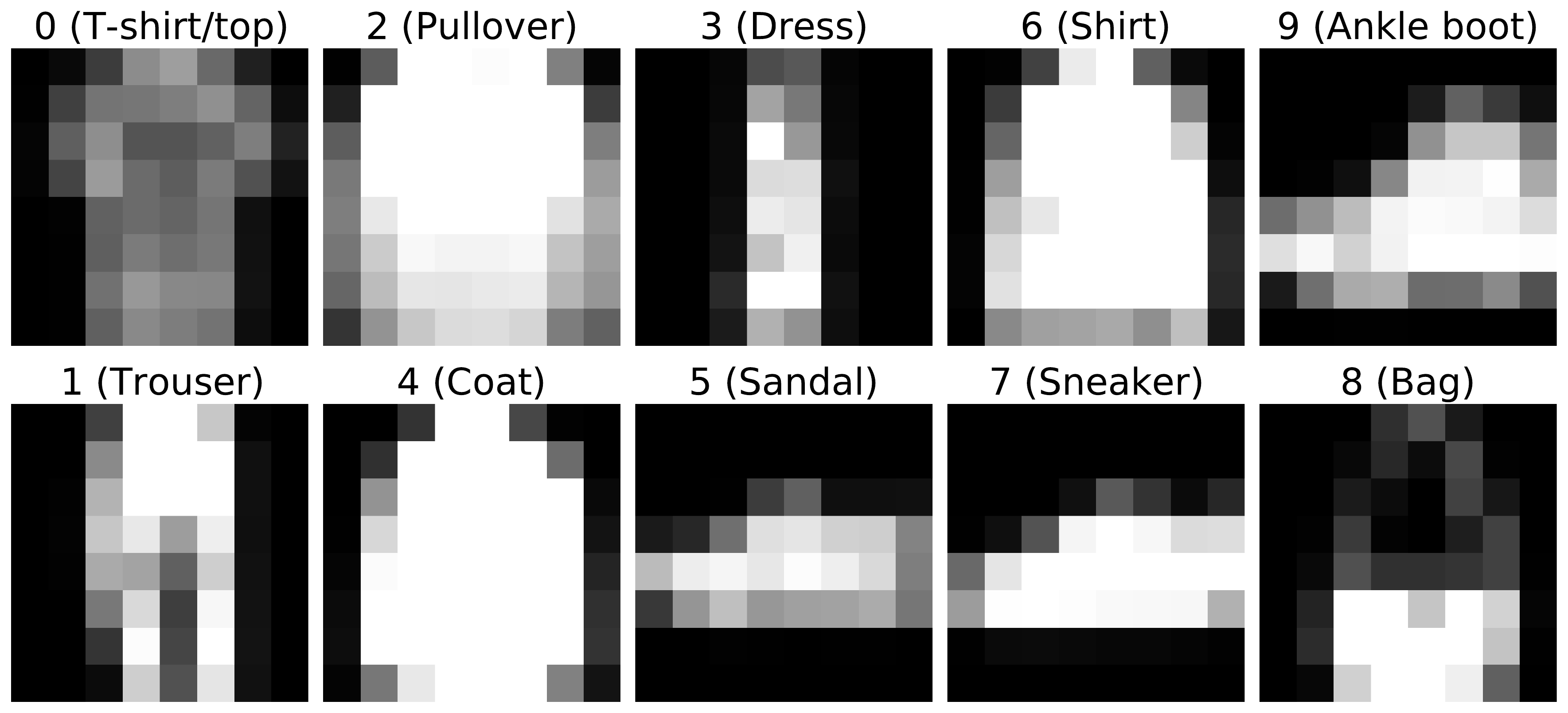}\label{fig:3}}
  \captionsetup{justification=raggedright, singlelinecheck=false}
  \caption{Example images of each original class in the three datasets, with the class label shown above each example. In each dataset, the classes in the top row are grouped into one and the classes in the bottom row are grouped into another for binary classification.}\label{fig:datasets}
\end{figure}

Samples from the three datasets used here are illustrated in Fig.~\ref{fig:datasets}. Compression of the images to dimension $8\times8$ allows tractable computation and optimization when ancillas are added to the tensor network QML models. Each pixel of an image is featurized through Eq.~\eqref{eq:qubit_encoding}. The three datasets have different levels of difficulty in terms of binary classification of grouped classes, with the MNIST dataset being the easiest one while the Fashion-MNIST dataset being the most challenging.

For each dataset, the numbers of training validation, and testing samples are $50040$, $9960$, and $10000$, respectively. The batch size used for training each model is $250$. We find that initializing the Hermitian matrices around zero, or equivalently the unitaries around the identity, yields better model performance. We use random normal distributions to initialize the entries (both the real and imaginary parts) of the Hermitian matrices, with means set to $0$ and standard deviation values
 tabulated below.  Tab.~\ref{tab:ttn_exps} corresponds to the experiments in Fig.~\ref{fig:three_datasets}, Tab.~\ref{tab:ttn_deph_net_only} corresponds to the experiments in Fig.~\ref{fig:deph_net_only}, and Tab.~\ref{tab:mera_exp} corresponds to the experiment in Fig.~\ref{fig:mera_mnist8pca}. The learning rates of the Adam optimizer are also tabulated respectively below. For each experiment, both the initialization standard deviation and learning rate are tuned with the help of Ray Tune~\cite{RayTune}.

\begin{table}[h]
    \caption{Standard deviations (std) of the initial values of the Hermitian matrices defining the TTN unitary matrices, and learning rates (lr) for the experiments of Fig.~\ref{fig:three_datasets} classifying each of the three datasets with unitary TTNs for which all qubits are subject to local dephasing.}\label{tab:ttn_exps}
    \begin{subtable}{\linewidth}
      \centering
        \caption{Compressed MNIST}
        \begin{tabular}{c|ccccccc}
            (std, lr) & $p$=0.0 & 0.1 & 0.2 & 0.4 & 0.6 & 0.8 & 1.0 \\
            \hline
            0 Anc. & (0.05, 0.005) & (0.05, 0.005) & (0.08, 0.005) & (0.1, 0.005) & (0.2, 0.005) & (0.3, 0.005) & (0.5, 0.005)\\
            
            1 Anc. & (0.07, 0.005) & (0.05, 0.005) & (0.08, 0.005) & (0.09, 0.005) & (0.1, 0.015) & (0.1, 0.015) & (0.1, 0.015)\\

            2 Anc. & (0.05, 0.005) & (0.05, 0.005) & (0.04, 0.005) & (0.03, 0.005) & (0.03, 0.005) & (0.02, 0.005) & (0.01, 0.005)\\

            3 Anc. & (0.05, 0.005) & --- & --- & --- & --- & --- & (0.01, 0.015)
        \end{tabular}
    \end{subtable}%
    \\
    \begin{subtable}{\linewidth}
      \centering
        \caption{Compressed KMNIST}
        \begin{tabular}{c|ccccccc}
            (std, lr) & $p$=0.0 & 0.1 & 0.2 & 0.4 & 0.6 & 0.8 & 1.0 \\
            \hline
            0 Anc. & (0.03, 0.005) & (0.03, 0.005) & (0.02, 0.005) & (0.01, 0.005) & (0.05, 0.005) & (0.01, 0.005) & (0.005, 0.005)\\
            
            1 Anc. & (0.03, 0.005) & (0.03, 0.005) & (0.05, 0.005) & (0.1, 0.005) & (0.15, 0.005) & (0.2, 0.005) & (0.3, 0.005)\\

            2 Anc. & (0.05, 0.005) & (0.05, 0.005) & (0.03, 0.005) & (0.01, 0.005) & (0.007, 0.005) & (0.007, 0.005) & (0.005, 0.005)\\

            3 Anc. & (0.05, 0.005) & --- & --- & --- & --- & --- & (0.01, 0.015)
        \end{tabular}
    \end{subtable} 
    \\
    \begin{subtable}{\linewidth}
      \centering
        \caption{Compressed Fashion-MNIST}
        \begin{tabular}{c|ccccccc}
            (std, lr) & $p$=0.0 & 0.1 & 0.2 & 0.4 & 0.6 & 0.8 & 1.0 \\
            \hline
            0 Anc. & (0.05, 0.005) & (0.05, 0.005) & (0.1, 0.005) & (0.2, 0.005) & (0.3, 0.015) & (0.4, 0.015) & (0.5, 0.015)\\
            
            1 Anc. & (0.5, 0.005) & (0.5, 0.005) & (0.3, 0.005) & (0.1, 0.005) & (0.05, 0.015) & (0.01, 0.015) & (0.005, 0.015)\\

            2 Anc. & (0.005, 0.005) & (0.005, 0.005) & (0.005, 0.005) & (0.005, 0.005) & (0.005, 0.015) & (0.005, 0.015) & (0.005, 0.015)\\

            3 Anc. & (0.005, 0.005) & --- & --- & --- & --- & --- & (0.05, 0.015)
        \end{tabular}
    \end{subtable} 
    \end{table}

    \begin{table}
      \centering
        \caption{Standard deviations (std) of the initial values of the Hermitian matrices defining the TTN unitary matrices, and learning rates (lr) for the experiments of  Fig.~\ref{fig:deph_net_only} classifying the compressed Fashion-MNIST dataset with  unitary TTNs for which all qubits are subject to local dephasing. }\label{tab:ttn_deph_net_only}
        \begin{tabular}{c|ccccccc}
            (std, lr) & $p$=0.0 & 0.1 & 0.2 & 0.4 & 0.6 & 0.8 & 1.0 \\
            \hline
            0 Anc. & (0.05, 0.005) & (0.05, 0.005) & (0.05, 0.005) & (0.04, 0.005) & (0.04, 0.005) & (0.03, 0.005) & (0.03, 0.005)\\
            
            1 Anc. & (0.5, 0.005) & (0.5, 0.005) & (0.4, 0.005) & (0.3, 0.005) & (0.2, 0.005) & (0.2, 0.005) & (0.1, 0.005)\\

            2 Anc. & (0.005, 0.005) & (0.005, 0.005) & (0.01, 0.005) & (0.03, 0.005) & (0.05, 0.005) & (0.06, 0.005) & (0.07, 0.005)\\

            3 Anc. & (0.005, 0.005) & --- & --- & --- & --- & --- & (0.05, 0.015)
        \end{tabular}
    \end{table}

    \begin{table}
      \centering
        \caption{Standard deviations (std) of the initial values of the Hermitian matrices defining the MERA unitary matrices, and learning rates (lr) for the experiments of Fig.~\ref{fig:mera_mnist8pca} 
        classifying the eight principle components of non-compressed MNIST images with 1D MERAs with all qubits subject to local dephasing.}\label{tab:mera_exp}
        \begin{tabular}{c|ccccccc}
            (std, lr) & $p$=0.0 & 0.1 & 0.2 & 0.4 & 0.6 & 0.8 & 1.0 \\
            \hline
            0 Anc. & (0.5, 0.005) & (0.4, 0.005) & (0.3, 0.005) & (0.2, 0.005) & (0.1, 0.005) & (0.07, 0.005) & (0.07, 0.005)\\
            
            1 Anc. & (0.3, 0.025) & (0.3, 0.025) & (0.3, 0.025) & (0.4, 0.025) & (0.4, 0.025) & (0.5, 0.025) & (0.5, 0.025)\\

        \end{tabular}
\end{table}

\end{appendix}
\end{document}